\documentclass[11pt]{article}

\usepackage{amssymb,amsmath,amsfonts}
\usepackage{latexsym}
\usepackage[totalwidth=17cm,totalheight=23.8cm]{geometry}

\usepackage{feynmp}
\usepackage[force]{feynmp-auto}
\usepackage{graphicx}

\newcommand{\R}{{\mathbb R}}

\newcommand{\im}{{\rm i }}

\newcommand{\be}{\begin{eqnarray}}
\newcommand{\ee}{\end{eqnarray}}

\let\a=\alpha

  \let\eps=\varepsilon

\let\==\equiv

\newcommand{\I}{\mathrm{i}}

\newcommand{\ket}[2]{ {\langle #1\,#2\rangle} }
\newcommand{\bra}[2]{ {[#1\,#2]}}

\newcommand{\cA}{\mathcal{A}}

\begin{document}

 \unitlength = 1mm
 
 \begin{fmffile}{defym}

\title{Deformations of Yang-Mills theory}
\author{Marco Cofano, Chih-Hao Fu and Kirill Krasnov \\ \it{School of Mathematical Sciences, University of Nottingham}\\ \it{University Park, Nottingham, NG7 2RD, UK}}

\date{v3: April 2015}
\maketitle

\begin{abstract} We introduce and study a new class of power-counting non-renormalisable gauge theories in four space-time dimensions. The Lagrangian is an arbitrary function of the self-dual part of the field strength. The resulting perturbation theory has the property that whenever two derivatives act on an internal line propagator, the result is a delta-function and the line collapses to a point. This means that there remains at most one derivative on each internal line, which gives improved ulta-violet behaviour. For many purposes, this class of theories behaves just like ordinary Yang-Mills theory. In particular, they all share the Yang-Mills theory MHV amplitudes. Moreover, these theories remain constructible (in the generalised sense), with the higher-point tree level scattering amplitudes obtainable from the lower-point amplitudes using the BCFW recursion relations, and adding new amplitudes at every particle number. Also, the square of these gauge-theory amplitudes gives the scattering amplitudes of ``deformation'' of General Relativity, at least for the low particle numbers that we checked. We compute the one-loop beta-function of the first new coupling constant, and find it to be positive, which signals the associated non-renormalisable interaction becoming important in the ulta-violet. 
\end{abstract}

\section{Introduction} 

By now it is commonly appreciated that every field theory is renormalisable provided all terms compatible with the symmetries are included into the Lagrangian; see \cite{Weinberg:2009bg} for a recent authoritative account of effective field theory. Moreover, it is in practice unnecessary to include all possible terms, as many of the terms can be eliminated by field redefinitions that do not change the physics, see e.g. \cite{Einhorn:2001kj} for a discussion of this in the context of the renormalisation group flow. 

This (and related \cite{Krasnov:2015kva}) paper can be viewed as an attempt to construct a renormalisable effective field theory model where only a very restricted type of terms is allowed to appear in the Lagrangian. Thus, we introduce and study a model where the number of terms in the Lagrangian is infinite, but much fewer as compared to what would be allowed in the full effective field theory. Our hope is that the class of theories we consider is still large enough to be closed under renormalisation. There is a simple mechanism, to be described below, that gives a justification for our (possibly over-optimistic) expectations, but beyond one loop we do not know whether our hope is realised. At one loop a definite statement can be made; see below.  

Thus, the purpose of this paper is to introduce and study a new infinite-parametric class of power-counting non-renormalisable theories. What is new and striking about these theories is that for many purposes they behave like the renormalisable theory (mother theory) from which they originated. The mother theory here is Yang-Mills theory in four space-time dimensions, and so we will refer to the new theories as {\it deformations} of Yang-Mills (YM) theory. 

\subsection{The class of theories}

The new theories are obtained by augmenting YM with certain new, importantly chiral, see below, power counting non-renormalisable interactions. The zeroth order idea is to consider effective Lagrangians of the form
\be\label{L-1}
{\cal L} = - \frac{1}{4g^2} F^2 + \frac{\alpha}{M^2} F^3 + \frac{\beta}{M^4} F^4 + \ldots,
\ee
where the first term is the usual YM Lagrangian, and the new terms are gauge-invariant scalars constructed from higher powers of the curvature. Note that there can be more than one new coupling constant at each order in the curvature, and we wrote the Lagrangian just schematically. Clearly, the coefficients in front of the new terms are dimensionful, and it is convenient to represent them as dimensionless couplings $\alpha,\beta,\ldots$ times an appropriate inverse power of some mass scale $M$. Alternatively, the Lagrangians of interest can be written as
\be\label{L-2}
{\cal L} = M^4 f(F/M^2),
\ee
where $f$ is a gauge- and Lorentz-invariant function of its curvature argument $F=F^a_{\mu\nu}$, where $a$ is the Lie algebra index. 

It is important to emphasise that the class of theories that we would like to consider is much smaller than the class of general effective field theories for the gauge field. Indeed, the effective field theory Lagrangian would also contain terms involving derivatives of the curvature. Thus, another gauge-covariant object that can be used in the construction of Lagrangians is the Lie-algebra valued vector$X_\mu^a:=d^\nu F_{\mu\nu}^a$. Building blocks with even larger numbers of derivatives are possible, e.g. $d^\mu d_\mu F_{\rho\sigma}^a$. It is clear that Lagrangians involving powers of $X_\mu^a$ would lead to field equations containing more than second derivatives of the gauge field. In contrast, the Lagrangians (\ref{L-2}) lead to second order in derivative field equations. It is clear that some of the higher-derivative terms can be eliminated by field redefinitions. But presumably there are terms that cannot be disposed off this way, and so (\ref{L-2}) is too restrictive as an effective field theory Lagrangian. 

Lagrangians of the type (\ref{L-2}) do appear as effective ones in a variety of theories. For example, one can imagine some fermionic fields interacting with the gauge field being integrated out. The resulting one loop effective action is non-local, but admits an expansion in powers of the field strength, with the scale $M$ being the typical mass of the particles integrated out. The famous example of an effective Lagrangian appearing this way is due to Euler-Heisenberg \cite{Heisenberg:1935qt}. Of course, such effective Lagrangians typically contain also the derivatives of the field strength, with (\ref{L-1}) being just the constant field approximation. 

Since the Lagrangian (\ref{L-2}) is non-renormalisable but contains much fewer terms than the effective field theory Lagrangian, one should expect that counterterms needed to cancel the arising loop divergences are not necessarily of the type already included in (\ref{L-2}). This is indeed the case. In other words, having added some non-renormalisable interactions one cannot stop and has to add all of them. The result is the effective field theory Lagrangian containing higher derivatives. This is not the route we want to follow in this paper. Instead, we will be studying theories of type (\ref{L-2}) with a certain further restriction added. 

We will further cut the class (\ref{L-2}) as follows. We have already restricted the theories of interest as compared to the general effective field theory by requiring the field equations to be second order in derivatives. This requires the Lagrangian to be a function of the field strength. We now further restrict what type of functions are allowed by requiring that all of our theories must share the same instanton sector. Recall that YM anti-self-dual (ASD) instantons are field configurations that have vanishing self-dual (SD) part of the curvature tensor $F^a_{\mu\nu}$, with $a$ being the Lie algebra index. It is convenient to describe the self-dual/anti-self-dual decomposition of the curvature in terms of spinor notations. In our conventions
\be
F^a_{MM'NN'} = \frac{1}{2} F^a_{MN} \epsilon_{M'N'} + \frac{1}{2} F^a_{M'N'} \epsilon_{NM}.
\ee
Here $F^a_{MM'NN'}$ is the spinor form of the curvature where each space-time index is replaced by a pair of spinor indices, and $F^a_{MN} = F^a_{MM'N}{}^{M'}, F^a_{M'N'}= F^{aM}{}_{M'MN'}$ are the self- and anti-self-dual parts of the curvature. Instantons, or ASD gauge fields are then field configurations satisfying 
\be\label{inst}
F^a_{MN}=0.
\ee
It is not hard to see that they are automatically solutions of the Yang-Mills equations. Indeed, as is well known, modulo a surface term, the YM Lagrangian can be rewritten as 
\be
{\cal L}_{YM} = -\frac{1}{4g^2} (F^a_{MN})^2.
\ee
The field equations following from this Lagrangian are
\be\label{YM-feqs}
d_{N'}{}^M F^a_{MN} = 0.
\ee
It is clear that the ASD gauge fields satisfying (\ref{inst}) are also solutions of (\ref{YM-feqs}). 

We now require this to continue to be true for an arbitrary member of our family of theories. It is easy to see that this implies changing (\ref{L-2}) into
\be\label{L*}
{\cal L} = M^4 f( F^a_{MN}/M^2),
\ee
i.e. into a function of only the self-dual part of the field strength. The resulting field equations are then
\be\label{feqs*}
d_{N'M}\left( \frac{\partial f}{\partial F^a_{MN}}\right) =0.
\ee
It is clear that on instantons (\ref{inst}) the matrix of first derivatives of $f$ appearing in the field equations is just a constant (possibly zero), and thus all field configurations (\ref{inst}) are also solutions of the field equations (\ref{feqs*}). 

One of the main goals of the present work is to characterise the class of theories (\ref{L*}) in as much detail as possible. In particular, we shall soon see that there is a simple mechanism that makes these theories behave like the usual Yang-Mills theory, in spite of a clear presence of power-counting non-renormalisable interactions. There are many other nice properties that theories (\ref{L*}) have, and this is what motivates our interest in them. 

\subsection{A 2-parameter example}

As a concrete example of a theory of the type (\ref{L*}) let us consider a gauge theory containing just a single additional interaction term as compared to the standard YM
\be\label{L2}
{\cal L} = - \frac{1}{4g^2} (F^a_{MN})^2 + \frac{\alpha}{3! g^2 M^2} f^{abc} F^a_A{}^B F^b_B{}^C F^c_C{}^A.
\ee
When a perturbative expansion is performed and the coupling constant is absorbed into the connection, one obtains $\alpha g/M^2$ in front of the new term, which is a convenient parameterisation of the new interaction, as we shall see. 

An effective field theory model of type (\ref{L2}) has been considered in the literature before; see in particular \cite{Dixon:2004za} for a study in the context of amplitudes. However, for reasons of unitarity, see below, the new interaction in (\ref{L2}) was always considered with its parity-dual. It is important to stress that we only allow in the Lagrangian the chiral half of this $F^3$-type interaction.  

\subsection{A summary of properties}

Here we give a brief account of the properties of (\ref{L*}), with some of these properties derived below, and some others to be obtained in \cite{Krasnov:2015kva}. Our main aim here is to indicate how nicely these theories behave, and thus hopefully stimulate further interest. 

Our main interest in (\ref{L*}) is due to the fact that, in spite of seemingly being a too small class of theories to remain closed under renormalisation, there is a simple mechanism that renders the divergences of these non-renormalisable theories essentially those of their renormalisable mother theory. To describe this mechanism, let us consider the perturbative expansion of Lagrangians (\ref{L*}). It is clear from (\ref{L2}) that the gauge field kinetic term is unchanged from what it is in YM theory. So, our theories describe interacting gluons, with some additional chiral interactions added on top of what is present in YM. An important observation is that the derivative of the gauge field enters the Lagrangian only in a very special combination 
\be\label{da}
(\partial a)_{AB}^a:=2 \partial_{(MM'} A^a_{N)}{}^{M'}. 
\ee
Unlike in YM theory, there are now interaction vertices with as many derivatives as the valency of the vertex. However, and this is important, there is at most just a single derivative acting on every leg of the vertex. So, at first sight our theories appear to diverge much worse than YM. However, there is a special property that when two derivatives happen to act on the same internal line propagator, the resulting second derivative is a multiple of the box operator, and then everything gets replaced by the $\delta$-function, see (\ref{ident}) for a graphical representation. As the result, many internal lines in Feynman diagrams of our theory will collapse to form effective higher-valent vertices. After the collapse, only internal lines with at most one derivative of the propagator remain. It can be seen that this is also what happens in Yang-Mills theory -- after some cancellations only diagrams with at most a single derivative on an internal line remain. Given that the number of derivatives on internal lines is what determines how diagrams diverge, we conclude that the divergences of our class of theories are essentially the same as in YM. See below for a further discussion of this, and also for the associated power-counting. This is the main justification for our hopes that the class (\ref{L*}), while clearly much smaller than the full effective field theory, may still be large enough to be closed under renormalisation.

This property of our theories is at work at any loop order, but we have not yet carefully analysed its consequences beyond the one loop. At one loop order there is a definite statement that can be made: The class of theories (\ref{L*}) is one-loop renormalisable. Thus, in spite of being clearly very far from the general effective field theory for the gauge field, no new one loop counter terms that are not already contained in (\ref{L*}) are required. All one loop divergences can be absorbed into either field redefinitions or coupling constant renormalisations. There will be  plenty of illustrations of why this is to be expected in the present paper, but the detailed proof of this statement is spelled out in \cite{Krasnov:2015kva}. 

Another important set of properties concerns the scattering amplitudes for (\ref{L*}). It is by now a well-known story that Yang-Mills theory is constructible, in the sense that all gluon scattering amplitudes can be determined by certain recursion relations from the basic 3-gluon ones. This means that all scattering amplitudes can be determined without recourse to any Lagrangian or Feynman rules, simply by constructing higher particle number amplitudes from lower ones using the recursion. In this paper we shall show that the same statement is true (in a generalised sense) for theories (\ref{L*}). One only has constructibility in a generalised sense because in conventional constructible theories like YM or GR there is only one coupling constant. As a result, one only needs the 3-point amplitudes as the seed for recursion. In our case, there is an infinite number of couplings. As we shall see, at the level of recursion relations these new coupling constants will enter at each particle number. Thus, at each gluon number a certain set of new amplitudes need to be added with new arbitrary couplings. After this is done one continues to determine higher point amplitudes by recursion as usual. Because of this, one could have discovered the class (\ref{L*}) by working solely with the scattering amplitudes, without recourse to any Lagrangian, as is the case with Yang-Mills theory. 

Our other remark is that such constructibility is certainly not the property of a general effective field theory for the gauge field. Such a theory contains higher derivative interactions, and this prevents the amplitudes from having strong enough falloff properties under the so-called Britto-Cachazo-Feng-Witten (BCFW) shifts \cite{Britto:2005fq}. As the result, the BCFW recursion does not work for amplitudes in an effective field theory of the gauge field, at least not directly. The (generalised) constructibility of (\ref{L*}), in addition to their nice behaviour under renormalisation, is our other main motivation for their study. 

Yet another motivation for considering the class of theories (\ref{L*}) is their relation to gravity. In \cite{Krasnov:2006du} a certain infinite-parametric family of ``deformations'' of General Relativity was introduced by one of the present authors. In \cite{Krasnov:2011up} they were recast as a family of diffeomorphism invariant SO(3) gauge theories, with GR being a member of this family \cite{Krasnov:2011pp}. In \cite{Delfino:2012aj} some graviton scattering amplitudes were computed using this ``pure connection'' formalism. In this paper we shall see that deformations of General Relativity \cite{Krasnov:2011up} continue to exhibit the 
\be
{\rm Gravity} =({\rm Gauge Theory})^2
\ee
pattern known to relate GR and YM, with appropriate squares of the deformed YM scattering amplitudes giving deformed gravity amplitudes. The fact that the double copy structure of gravity also extends to (some) higher order terms in the gauge theory and gravity Lagrangians is not new; it has been observed and studied in particular in \cite{Broedel:2012rc}. Thus, our observations in this regard confirm the pattern noticed earlier. 

\subsection{Unitarity and Interpretation}

Our final introductory remark will likely make some readers less interested in the whole story, but it is difficult to avoid mentioning from the outset. In Euclidean (or split) signature the self-dual part of the field strength is real. So, at the very least the theories we study make sense as Euclidean (split) signature field theories with intriguing properties. 

In Lorentzian signature the self-dual part of the field strength is complex. This makes the Lagrangians (\ref{L*}) complex. This immediately raises the question about unitarity and more general interpretation of the resulting quantum theory. In the next section we will give a quick Hamiltonian analysis of (\ref{L*}), and see that the corresponding Hamiltonians are not Hermitian, but instead parity-time $(PT)$-symmetric. Hence one is in the domain developed by the proponents of the $PT$-symmetric quantum mechanics \cite{Bender:2007nj}. As is explained in e.g. \cite{Mostafazadeh:2010yx}, one way to give physical interpretation to such systems is to find a new positive definite inner product that makes the Hamiltonian a Hermitian operator. As the study of concrete $PT$-symmetric quantum system shows, see e.g. \cite{Mostafazadeh:2004qh}, the equivalent Hermitian Hamiltonian can be expected to be extremely complicated, and so in practice for many purposes, e.g. determination of the energy levels, it is best to study the original non-Hermitian system. However, the non-trivial positive definite inner product that makes the Hamiltonian Hermitian plays an important role in the physical interpretation of the $PT$-symmetric quantum system. 

In this paper we make no attempt to determine (or establish existence of) the inner product that makes our quantum theory unitary. We shall study it as is, working with the obvious inner product in which the S-matrix is not unitary. The corresponding unitary S-matrix is then to be found from the non-unitary one by an appropriate similarity transformation. Alternatively, our scattering amplitude calculations can be interpreted as those for the split signature, where everything is real, but no physical interpretation is available. 

Thus, we make no claim that the quantities computed in this paper are ones that can be compared to some experimental measurements. Our interest in the class of theories (\ref{L*}) is not because we want to use one of such Lagrangians to describe what happens in an experiment, at least not yet. For now we are interested in these theories because of the striking properties that they exhibit. Thus, our main goal here is to characterise the theories (\ref{L*}) in as much details as possible. Determination of an appropriate physical interpretation of (\ref{L*}), if any, is only possible when their properties are well understood, and so is left to the future. 

In addition to non-unitarity discussed above, our theories are also plagued by more conventional unitarity problems arising in any non-renormalisable theory. This is the tree-level unitarity violation arising because in theories with negative mass dimension couplings the tree level scattering amplitudes at sufficiently high energies become larger than unity. This problem can only be cured (if at all) by an ultraviolet completion of the theory. The main objective of this paper can be paraphrased as to make a step in the direction of finding this UV completion for (\ref{L*}), by studying the arising renormalisation group flow and hoping for a UV completion in the sense of asymptotic safety of Weinberg \cite{Weinberg:2009bg}.

\subsection{Organisation of the paper}

We start in section \ref{sec:Ham} with the Hamiltonian analysis of our theories. This illustrates what kind of complex Lagrangians we deal with. We also state the mode decomposition for the gauge field here, as well as fix our conventions, in particular for the helicity spinors. Section \ref{sec:Feyn} then describes two different gauge-fixings, one useful for loop calculations and the other useful for colour-ordered Feynman rules. We also derive the central ``collapsing'' property of our perturbation theory here, and evaluate the cubic vertices on-shell. In section \ref{sec:Recur} we discuss how BCFW recursion relations work for our theories. The main point here is that the amplitudes with at least one negative and one positive helicity gluons continue to be determinable by the usual BCWF recursion. The exception is the amplitudes with all plus helicity gluons. For these we need a more involved shift on all momenta. Section \ref{sec:4-point} uses the BCFW recursions to compute the 4-gluon scattering amplitudes, and section \ref{sec:double} discusses their double copy structure. In section \ref{sec:renorm} we analyse the power counting for our theories, and illustrate it on an example of a divergent self-energy diagram. Section \ref{sec:beta} computes the triangle diagrams necessary to extract the beta-function of the new coupling. We conclude with a discussion. 

\section{Hamiltonian formulation and the mode decomposition}
\label{sec:Ham}

In this section we perform a quick Hamiltonian analysis of our theories. One justification for this exercise is to see what kind of complex Hamiltonians result. Another justification is to fix conventions and then state the mode decomposition of the gauge field. In the next section we use this mode decomposition to obtain the Feynman rules for computing the S-matrix. 

\subsection{First order formulation}

To perform the Hamiltonian analysis of (\ref{L*}) it is most useful to start with an equivalent but first order in derivatives formulation. Introducing an auxiliary field $B^a_{MN}$ we write
\be\label{1-order}
{\cal L} = B^{a MN} F^a_{MN} - M^4 V(B/M^2),
\ee
where $V(B)$ is a gauge- and Lorentz- invariant function of the auxiliary field. The field equation for the auxiliary field is $F^a_{MN} = M^2 (V')^a_{MN}$, where $V'$ is the matrix of first derivatives of $V$. It is clear that $V(B/M^2)$ is simply the Legendre transform of $f(F/M^2)$. 

\subsection{Hamiltonian formulation}

We now perform the $3+1$ decomposition of (\ref{1-order}). In Lorentzian signature the relevant formulas are
\be
F^a_{MN} = (2\im F^a_{0i} - \epsilon^{ijk} F^a_{jk}) \frac{1}{\sqrt{2}} T^i_{MN},
\ee
where $i,j,k$ are the spatial indices and $T^i_{MN}$ are multiples of Pauli matrices fixed by the following algebra they satisfy
\be\label{T}
T^{iAB} T^j_B{}^C = -\frac{1}{2} \delta^{ij} \epsilon^{AC} + \frac{1}{\sqrt{2}} \epsilon^{ijk} T^{k AC}.
\ee
We also have
\be
F^a_{0i}=\dot{A}^a_i - d_i A^a_0,
\ee
where $d_i$ is the covariant derivative with respect to the spatial connection. Introducing the new auxiliary field $P^a_i$ so that
\be
B^a_{MN} = \frac{1}{\im\sqrt{2}} P^{ai} T^i_{MN}
\ee
we can rewrite the Lagrangian as
\be
{\cal L}= P^{ai} \dot{A}^a_i + A^a_0 d_i P^{ai} - {\cal H} ,
\ee
where
\be
H^{ai} = \frac{1}{2} \epsilon_{ijk} F^a_{jk}
\ee
is the magnetic field and the Hamiltonian is
\be
{\cal H} = -\im P^{ai} H^a_i + M^4 V(P/M^2).
\ee
We now see that the field $P^a_i$ is the canonically conjugate momentum to the spatial gauge field$A_i^a$.

For Yang-Mills theory we have
\be
M^4 V_{\rm YM}(B/M^2)= - g^2 (B^a_{MN})^2 = \frac{g^2}{2} (P^{ai})^2.
\ee
Thus, in this case we can rewrite the Hamiltonian as
\be
{\cal H}_{\rm YM} = \frac{g^2}{2}\left( P^{ai} - \frac{\im}{g^2} H^{ai}\right)^2 + \frac{1}{2g^2} (H^{ai})^2.
\ee
We recognise the canonical transformation of the usual Yang-Mills Hamiltonian. It can be checked that the above shift of the momentum variable is indeed a legitimate canonical transformation in that the symplectic form is unchanged. Thus, the correct reality conditions in this case are that the spatial connection, and thus the magnetic field are real, and the combination 
\be\label{EP}
E^{ai} := P^{ai} - \frac{\im}{g^2} H^{ai}
\ee
is real. This renders the Hamiltonian real, and theory unitary.

For the theory (\ref{L-2}) deformed by a cubic term we get
\be\label{Ha}
{\cal H} = \frac{g^2}{2}\left( E^{ai} \right)^2 + \frac{1}{2g^2} (H^{ai})^2 - \frac{\im\alpha g^4}{3M^2} f^{abc} \epsilon^{ijk} \left( E^a_i + \frac{\im}{g^2} H^a_i\right)\left( E^b_j + \frac{\im}{g^2} H^b_j\right)\left( E^c_k + \frac{\im}{g^2} H^c_k\right)
\ee
plus higher order terms. Thus, if we require $E$ and $H$ to be real, the Hamiltonian is not Hermitian. However, if $\alpha\in\R$ the Hamiltonian (\ref{Ha}) does exhibit an anti-linear discrete symmetry 
\be\label{P}
E^{ai}\to - E^{ai}, \qquad H^{ai}\to H^{ai}
\ee
this followed by the complex conjugation. It is clear that this is the case for any $V(B/M^2)$, or equivalently $f(F/M^2)$ that contains only real coefficients when expanded into powers of its argument. In what follows we shall assume that the functions $f,V$ satisfy this requirement. 

A remark about (\ref{Ha}) is that this way of writing the effect of deformation makes it clear that at least the cubic deformation is only possible for the non-abelian Yang-Mills theory. Indeed, in the case of Maxwell we would not have an additional Lie-algebra index on the fields $E_i,H_i$, and the vector product present in (\ref{Ha}) would simply be zero. So, for a ${\rm U}(1)$ gauge field the simplest new interaction is quartic in the field strength.

\subsection{Pseudo-Hermitian Hamiltonians}

As is explained in the literature on $PT$-symmetric quantum mechanics, see e.g. \cite{Mostafazadeh:2010yx}, it is in the presence of an anti-linear symmetry that one may hope to make sense of the arising quantum mechanical system. 

In practice, this requires finding a metric operator $\eta$ satisfying
\be
H^\dagger = \eta H \eta^{-1},
\ee
where the dagger is defined with respect to the "obvious" inner product, i.e. the one with respect to which the operator $H$ is not Hermitian. The operators for which such $\eta$ exists are called pseudo-Hermitian. One then defines the new inner product
\be
\langle \phi| \psi\rangle_\eta := \langle \phi | \eta \psi\rangle
\ee
with respect to which the Hamiltonian is Hermitian. If the new inner product happens to be positive-definite, it defines a new Hilbert space where the original Hamiltonian acts as a Hermitian operator. Mostafazadeh, see \cite{Mostafazadeh:2010yx} and references therein, has shown that a necessary and sufficient condition for an existence of $\eta$ for a Hamiltonian $H$ with a complete set of eigenvectors is that it commutes with an invertible anti-linear operator. 

In our case this operator is that given by (\ref{P}) followed by the complex conjugation. Thus, one may hope for an existence of an appropriate metric operator $\eta$. This would give one way of giving physical interpretation to our theories. However, as we have already explained in the Introduction, our main aim here is not to develop such a physical interpretation but instead study (formal for now) properties of theories (\ref{L*}). Thus, even though all calculations below are performed in Lorentzian signature and the arising S-matrix is not unitary in the obvious inner product, our treatment can be justified by going e.g. to the split signature. In this case there is still a scattering theory of massless particles that can be set up, but no unitarity to be concerned about. Of course our treatment of the scattering theory would be more compelling if we also found an inner product that makes everything unitary. But this is a difficult problem that likely requires a very different sort of considerations from the ones given below. We hope to return to this problem in the future. 

\subsection{Mode decomposition}

We finish this section by stating the mode decomposition for the gauge field. We first note that all new terms in the Lagrangian only contribute to the interactions, and the kinetic term for the gauge-field is unchanged from what it is in Yang-Mills theory. Thus, the free Hamiltonian continues to be given by
\be
{\cal H}_{\rm free} = -\im P^{ai}  \epsilon^{ijk} \partial_j A^a_k + \frac{g^2}{2} (P^{ai})^2, 
\ee
where we have only included into the Hamiltonian the part of the magnetic field linear in the gauge-field. The Hamilton-Jacobi equations can be written as
\be\label{HJ}
g^2 P^{ai} = \left(\frac{\partial}{\partial t} + \im \epsilon\partial \right) A^{ai},\qquad
\left(\frac{\partial}{\partial t} - \im \epsilon\partial \right) P^{ai} = 0,
\ee
where the operator $\epsilon\partial$ acting on any Lie-algebra valued spatial vector $X^{ai}$ is defined as
\be
(\epsilon\partial X)^{ai} := \epsilon^{ijk} \partial_j X^a_k.
\ee
It is an easy exercise to check
\be
(\epsilon\partial)^2 X^{ai} = \partial^i \partial_j X^{aj} -\partial^j\partial_j X^{ai}.
\ee
Thus, on transverse vectors it reduces to minus the Laplacian. Therefore, as expected, the two equations (\ref{HJ}) together imply that the transverse $\partial^i A^a_i=0$ part of the connection satisfies the wave equation.

We can now write the mode decomposition for the transverse part of $A^a_i$
\be\label{dec}
A^a_i = g \int \frac{d^3k}{(2\pi)^3 2\omega_k} \left( \epsilon^-_i (a^-_k)^a e^{\im kx} +  \epsilon^+_i (a^+_k)^a e^{\im kx} + \epsilon^+_i (a^-_k)^\dagger{}^a e^{-\im kx} +  \epsilon^-_i (a^+_k)^\dagger{}^a e^{-\im kx}\right).
\ee
The factor of $g$ in front will of course later be absorbed into the gauge field perturbation. Our signature is $-,+,+,+$ so that $kx=k^i x_i - \omega t$. For our massless gluons $\omega_k=|k|$. The vectors $\epsilon^\pm_i$ are the two circular polarisation helicities. They are chosen to satisfy
\be\label{op-eps}
\left(\frac{\partial}{\partial t} + \im \epsilon\partial \right) \epsilon^-_i  e^{\pm \im kx} = 0 \quad \Leftrightarrow \quad \im \epsilon_{i}{}^{mn} k_m \epsilon^-_n =\omega_k \epsilon^-_i
\ee
and
\be
\epsilon^+_i=(\epsilon^-_i)^*.
\ee
The condition (\ref{op-eps}) guarantees that only the positive polarisation gives a non-zero momentum vector $P^{ai}$. 

The decomposition (\ref{dec}) together with the equal time commutation relations $[A^a_i(x),P^b_j(y)]=\im\delta^{ab} \Pi_{ij} \delta^3(x-y)$, where $\Pi_{ij}$ is the transverse projector, implies the correct commutation relations
\be
[(a^\pm_k)^a, (a^\pm_p)^{\dagger b}] = (2\pi)^3 2\omega_k \delta^{ab} \delta^3(k-p)
\ee
provided the following normalisation condition is satisfied
\be
\epsilon^+_i \epsilon^-_j + \epsilon^-_i \epsilon^+_j = \Pi_{ij}.
\ee 
Thus, in a coordinate system in the momentum space in which the $z$-direction is taken to be along the direction of the momentum vector $k_i$, the polarisation vectors are given by
\be
\epsilon^\pm_i = \frac{1}{\sqrt{2}} (x_i \pm \im y_i).
\ee
In particular $\epsilon^+_i \epsilon^-_i = 1$. 

\subsection{Mode decomposition in the spinor form}

The last equation we need to set up the theory is the mode decomposition in the spinor form. Writing $A^a_{MN}=A^a_i T^i_{MN}$, where the matrices $T^i_{MN}$ have already been introduced above, and then following the rules for converting into a space-time form described in subsection 6.10 of \cite{Delfino:2012aj}, we get
\be\label{mode}
A^a_{MM'} = g \int \frac{d^3k}{(2\pi)^3 2\omega_k} \left( \epsilon^-_{MM'} (a^-_k)^a e^{\im kx} +  \epsilon^+_{MM'} (a^+_k)^a e^{\im kx} + \epsilon^+_{MM'} (a^-_k)^\dagger{}^a e^{-\im kx} +  \epsilon^-_{MM'} (a^+_k)^\dagger{}^a e^{-\im kx}\right)
\ee
with
\be\label{helicity}
\epsilon^-_{MM'} = \im \frac{q_M k_{M'}}{\ket{q}{k}},\qquad \epsilon^+_{MM'} = \im \frac{k_M q_{M'}}{\bra{q}{k}}.
\ee
Here $\ket{\lambda}{\mu} := \lambda^A \mu_A, \bra{\lambda}{\mu}:= \lambda_{A'}\mu^{A'}$ are the two spinor contractions, and $q_M, q_{M'}$ are the reference spinors whose presence reflects the gauge transformation freedom. It is easy to see that the required normalisation condition is satisfied $\epsilon^{+M}{}_{M'} \epsilon^-_M{}^{M'}=1$.

We now have the usual rules of the Lehmann-Symanzik-Zimmermann (LSZ) reduction, with the understanding that the helicity spinors (\ref{helicity}) must be inserted into the amputated Feynman diagrams to extract the scattering amplitudes. 

\section{Gauge-fixing and the Feynman rules}
\label{sec:Feyn}

There are two useful gauge-fixings that we will consider that lead to two different versions of the Feynman rules. One of them is more useful for one-loop calculations, the other is useful to derive the colour ordered Feynman rules. Since the new terms that we added to the Lagrangian only affect interactions, the gauge-fixing(s) that we need to introduce are the same as for YM theory. The only slight novelty here is that we work with a chiral formulation of YM, so even the resulting YM Feynman rules are slightly different from the usual ones. In particular, spinor notations will be crucial for writing all formulas.

\subsection{Feynman gauge}

Absorbing the YM coupling into the gauge field perturbation, the kinetic term for the Lagrangian (\ref{L2}) reads
\be\label{kin-1}
{\cal L}^{(2)} = -\frac{1}{2} (\partial_{MM'} A^a_{N}{}^{M'} + \partial_{NM'} A^a_{M}{}^{M'}) \partial^M{}_{N'} A^{a NN'}.
\ee
Using the magic of spinor identities we can rewrite the second term in brackets as
\be\label{i-1}
\partial_{NM'} A^a_{M}{}^{M'} = \partial_{MM'} A^a_{N}{}^{M'} + \epsilon_{MN} \partial^E{}_{M'} A^a_{E}{}^{M'}.
\ee
This formula is easily checked by multiplying both sides with $\epsilon^{MN}$. We also use the following rules for raising and lowering of the spinor indices $\epsilon^{MN} \lambda_N = \lambda^M, \lambda^M\epsilon_{MN} = \lambda_N$, and the same for the primed indices. Thus, we can rewrite (\ref{kin-1}) as
\be
{\cal L}^{(2)} = - \partial_{MM'} A^a_{N}{}^{M'} \partial^M{}_{N'} A^{a NN'} + \frac{1}{2} (\partial^\mu A^a_\mu)^2.
\ee
where our convention for the metric contraction is $u^\mu v_\mu = u^{M}{}_{M'} v_M{}^{M'}$. In the first term here there is a contraction of the partial derivatives in an unprimed spinor index. Given that the partial derivatives commute we have
\be
\partial_{MM'} \partial^M{}_{N'} = \frac{1}{2} \epsilon_{N'M'} \partial^\mu \partial_\mu.
\ee
This formula is easily checked by multiplying both sides with $\epsilon^{N'M'}$. Thus, overall, modulo surface terms
\be
{\cal L}^{(2)} = - \frac{1}{2} (\partial_\mu A^a_\nu)^2  + \frac{1}{2} (\partial^\mu A^a_\mu)^2,
\ee
as expected. 

We now add a gauge-fixing term that removes the last term, and add the ghost Lagrangian
\be
{\cal L}_{\rm ghost} = \bar{c}^a \partial^\mu d_{\mu} c^a,
\ee
where $d_\mu$ is the covariant derivative with respect to the gauge field, and $c^a, \bar{c}^a$ are the ghost and antighost fields. Overall the gauge-fixing term and the ghost Lagrangian are exactly as in the usual YM theory.

\subsection{The Feynman gauge Feynman rules}

The momentum space propagator for the gauge field follows directly from the gauge-fixed kinetic term and reads
\be
\label{propagator}
	\parbox{20mm}{\begin{fmfgraph*}(20,3) 
	  \fmftop{i,o}
	  \fmfv{label=$\mu i$,label.angle=90}{i}
	  \fmfv{label=$\nu j$,label.angle=90}{o}
	\fmf{wiggly}{i,o}
	   \end{fmfgraph*}}\,\, \qquad = \quad  \frac{1}{\I k^2}({-\eps_{MN}\eps_{M'N'}}) 
\ee
The extra minus sign here is due to our convention for the metric $\eta_{MM'NN'}=-\epsilon_{MN}\epsilon_{M'N'}$. 

To derive the vertices we first note that the derivative of the gauge field only appears in the combination (\ref{da}). For reasons that will become clear below, it is then convenient to represent the cubic YM vertex as a sum of three terms, where in each of the terms the derivative is applied to a different external leg. We denote the leg where the derivative acts with a dot. The resulting Feynman rule is then
\fmfcmd{%
style_def dotted expr p =
cdraw (wiggly p);
filldraw fullcircle scaled 5 shifted point length(p)/4 of p
enddef;}
\fmfcmd{%
style_def dottedf expr p =
cdraw (wiggly p);
filldraw fullcircle scaled 5 shifted point length(p)/2 of p
enddef;}
\be   \label{v3-YM}
  \parbox{20mm}{\begin{fmfgraph*}(20,20) 
  \fmftop{i}
  \fmfbottom{o1,o2}
\fmf{wiggly}{v,i}
  \fmf{wiggly}{v,o1}
  \fmf{wiggly}{v,o2}
   \end{fmfgraph*}}\,\,  \qquad = \quad \parbox{20mm}{\begin{fmfgraph*}(20,20) 
  \fmftop{i}
  \fmfbottom{o1,o2}
\fmf{dotted}{v,i}
  \fmf{wiggly}{v,o1}
  \fmf{wiggly}{v,o2}
   \end{fmfgraph*}}\,\,+\,\, \parbox{20mm}{\begin{fmfgraph*}(20,20) 
  \fmftop{i}
  \fmfbottom{o1,o2}
\fmf{wiggly}{v,i}
  \fmf{dotted}{v,o1}
  \fmf{wiggly}{v,o2}
   \end{fmfgraph*}}\,\,+\,\, \parbox{20mm}{\begin{fmfgraph*}(20,20) 
  \fmftop{i}
  \fmfbottom{o1,o2}
\fmf{wiggly}{v,i}
  \fmf{wiggly}{v,o1}
  \fmf{dotted}{v,o2}
   \end{fmfgraph*}}  
\ee
where
\be   
 \label{v3old}
 \parbox{20mm}{\begin{fmfgraph*}(20,20) 
  \fmftop{i}
  \fmfbottom{o1,o2}
  \fmfv{label=$q\; c$,label.angle=90}{o1}
  \fmfv{label=$p\; b$,label.angle=90}{o2}
  \fmfv{label=$k \;a$,label.angle=0}{i}
\fmf{dotted}{v,i}
  \fmf{wiggly}{v,o1}
  \fmf{wiggly}{v,o2}
   \end{fmfgraph*}}\,\,  \qquad = \quad 
    g f^{abc} (k A(k))^{aAB} A_{A B'}^b(p) A^{c}_{B}{}^{B'}(q)
\ee
We have defined 
\be
(k A(k))^{aAB} := 2 k^{(A}_{A'} A^{aB)A'}(k)
\ee
and assumed that all the momenta are incoming, and $A(k), A(p), A(q)$ are the placeholders. These can later be either replaced with external polarisations, or Wick contracted with another placeholder to form the propagator. 
    
The YM quartic vertex reads
\be
 \label{v4old}
\parbox{20mm}{\begin{fmfgraph*}(20,20) 
   \fmftop{i1,i2}
     \fmfbottom{o1,o2}
   \fmfv{label=$p\;e$,label.angle=180}{o1}
   \fmfv{label=$q\;d$,label.angle=0}{o2}
   \fmfv{label=$k\;a$,label.angle=180}{i1}
   \fmfv{label=$l\;b$,label.angle=0}{i2}
   \fmf{wiggly}{v,i1}
   \fmf{wiggly}{v,i2}
   \fmf{wiggly}{o1,v}
   \fmf{wiggly}{o2,v}
    \end{fmfgraph*}}\,\, \qquad = \quad  -2\I g^2 f^{abc}f^{cde}A_{(MM'}^a(k) A_{N)}^{b} {}^{M'}(l) A_{}^{d(M}{}_{N'}(a) A^{e N)N'}(p) + \text{2 terms} 
\ee
The additional terms that we did not write are analogous, just with different pairs of placeholder gauge fields contracted into the structure constants. 

The cubic and quartic vertices following from the new cubic term (second term in (\ref{L2})) are as follows. First, there is a completely symmetric cubic interaction
\be
 \label{v3new}
\parbox{20mm}{\begin{fmfgraph*}(20,20) 
   \fmftop{i}
     \fmfbottom{o1,o2}
   \fmfv{label=$q\;c$,label.angle=90}{o1}
   \fmfv{label=$p\;b$,label.angle=90}{o2}
   \fmfv{label=$k\;a$,label.angle=0}{i}
 \fmf{dotted}{v,i}
   \fmf{dotted}{v,o1}
   \fmf{dotted}{v,o2}
    \end{fmfgraph*}}\,\, \qquad = \quad   
        \frac{\a g}{M^2} f^{abc}(k A(k))^a_A{}^B (pA(p))^b_B{}^C (qA(q))^c_C{}^A \\
\ee
Then there is the following quartic vertex
\be
\parbox{20mm}{\begin{fmfgraph*}(20,20) 
   \fmftop{i1,i2}
     \fmfbottom{o1,o2}
   \fmfv{label=$p\;e$,label.angle=180}{o1}
   \fmfv{label=$q\;d$,label.angle=0}{o2}
   \fmfv{label=$k\;a$,label.angle=180}{i1}
   \fmfv{label=$l\;b$,label.angle=0}{i2}
   \fmf{dotted}{v,i1}
   \fmf{dotted}{v,i2}
   \fmf{wiggly}{o1,v}
   \fmf{wiggly}{o2,v}
    \end{fmfgraph*}}\,\, \qquad + \text{5 terms}
 \ee
 where
 \be   
 \label{v4new}
\parbox{20mm}{\begin{fmfgraph*}(20,20) 
   \fmftop{i1,i2}
     \fmfbottom{o1,o2}
   \fmfv{label=$p\;e$,label.angle=180}{o1}
   \fmfv{label=$q\;d$,label.angle=0}{o2}
   \fmfv{label=$k\;a$,label.angle=180}{i1}
   \fmfv{label=$l\;b$,label.angle=0}{i2}
   \fmf{dotted}{v,i1}
   \fmf{dotted}{v,i2}
   \fmf{wiggly}{o1,v}
   \fmf{wiggly}{o2,v}
    \end{fmfgraph*}}\,\,    \qquad = \quad   -2\I \frac{\a g^2}{M^2} f^{abc}f^{cde}(k A(k))_A^{a}{}^B (l A(l))_B^{b}{}^C A_{CM'}^{d}(q) A^{e A M'}(p) 
  \ee
The new term in (\ref{L2}) produces also  quintic and sextic interaction vertices, but we will not need them here. 

\subsection{Central identity}

The benefit of using the dot notation consists in the fact that when two dots appear on the same internal line, this line collapses to a point. Indeed, consider the following Wick contraction
\begin{align}  \label{2dotscontr}
	\left\langle \;(k A^a )_{AB}\; (k A^b)_{CD}\;\right\rangle &= 	\left\langle \; 2k_{(AA'} A^a_{B)}{}^{A'} \; 2 k_{(C C'} A^b_{D)}{}^{C'}\;\right\rangle  \nonumber \\
	&=- 4 \delta^{ab} k_{(AA'}\frac{1}{\I k^2}\eps^{A'C'}\eps_{B)(D}k_{C)C'} = \frac{2}{\I} \delta^{ab}\eps_{A(C} \eps_{|B|D)}.
\end{align}
We thus see that the $k^2$ in the denominator gets cancelled by $k^2$ arising in the numerator. The way this happens is instructive. To understand this we need to trace the contraction of the primed indices only. These are contracted in each of the two factors of $(kA)$. Then the contraction of the two placeholder $A$'s makes the primed indices of the two factors of $k$ contract. This produces a factor of $k^2$ times an $\epsilon$ of the unprimed indices. It is this $k^2$ appearing in the numerator that cancels the propagator. 

The identity (\ref{2dotscontr}) can be represented graphically as follows
\be\label{ident}
\parbox{20mm}{\begin{fmfgraph*}(20,20) 
   \fmfleft{i1,i2}
     \fmfright{o1,o2}
      \fmf{wiggly}{i1,i1p,v1,i2p,i2}
            \fmf{dots,left=0.5,tension=0.2}{i1p,i2p}
   \fmf{dottedf}{v1,v3}
   \fmf{dottedf}{v3,v2}
   \fmf{wiggly}{o1,o1p,v2,o2p,o2}
   \fmf{dots,right=0.5,tension=0.2}{o1p,o2p}
    \end{fmfgraph*}}\,\, \qquad = \qquad 2\im 
    \parbox{20mm}{\begin{fmfgraph*}(20,20) 
   \fmfleft{i1,i2}
     \fmfright{o1,o2}
      \fmf{wiggly}{i1,i1p,v1,i2p,i2}
            \fmf{dots,left=0.5,tension=0.2}{i1p,i2p}
     \fmf{wiggly}{o1,o1p,v1,o2p,o2}
   \fmf{dots,right=0.5,tension=0.2}{o1p,o2p}
    \end{fmfgraph*}}
  \ee
where an additional minus sign comes from the fact that on the internal line one of the momenta is incoming and the other is outgoing. The dots represent any lines that may or may not be present on the left and the right. The lines that are external in this diagram may also contain derivatives, i.e. other dots. Note that the effective vertex arising needs to be worked out on a case-by-case basis, see examples of this below.  

The identity (\ref{2dotscontr}) clearly holds due to the very special way (\ref{da}) that the derivatives of the gauge-field appear in the interaction vertices. Clearly, no such identity would hold if just the anti-symmetrised derivative $\partial_{[\mu} A^a_{\nu]}$ would be involved, as in the usual Yang-Mills Feynman rules. This is the benefit of the ``chiral'' formulation considered here. As we shall soon see, it is the identity (\ref{ident}) that is at the heart of the improved properties of our theories as compared to a general effective field theory. 

The identity (\ref{2dotscontr}) is not new; it has been noticed in particular in \cite{Boels:2008ef}, see (5.10) of this paper and in \cite{Boels:2008fc} see (4.15). The  ``dot'' notation introduced here, however, appears to be new. This notation is particularly useful for analysis of the arising diagrams, as well as their cancellations. Thus, using our notation it is for example very easy to verify that the only purpose of existence of the YM theory quartic vertex is to cancel the YM cubic vertex diagrams where the two dots happen to appear on the same internal line. Because of this fact, the usual YM theory can be recast into a theory with only cubic vertices (\ref{v3-YM}) as well as the rule that there is no more than one dot on any line. This fact appears to be important, and will be explored elsewhere. 

\subsection{Gervais-Neveu gauge}

To derive the colour-ordered Feynman rules convenient for the scattering amplitude calculations, we introduce a different gauge. We basically follow \cite{Srednicki:2007qs} Chapter 79, with the only difference being that spinor indices are used instead of space-time, and the original YM Lagrangian is chiral. 

Let us introduce the matrix-valued generators $T^a$ satisfying
\be
{\rm Tr}(T^a T^b)=\delta^{ab}, \qquad [T^a,T^b]=\im \sqrt{2} f^{abc} T^c.
\ee
We define the matrix-valued field
\be
A_{MM'} := T^a A^a_{MM'}.
\ee
The self-dual part of the matrix-valued curvature is then
\be\label{F-mat}
F_{MN} = 2\partial_{(MM'} A_{N)}{}^{M'} + \frac{g}{\im\sqrt{2}}(A_{MM'}A_N{}^{M'} + A_{NM'}A_M{}^{M'}),
\ee
and the chiral Yang-Mills Lagrangian is
\be
{\cal L}_{\rm YM} = -\frac{1}{4} {\rm Tr}( F^{MN} F_{MN}).
\ee

The gauge-fixing term is supposed to make the kinetic term for the gauge field non-degenerate. Previously this has been achieved by adding $-(1/2) (\partial^\mu A_\mu^a)^2$ to the Lagrangian. This only modifies the kinetic term, leaving the interaction vertices unchanged. However, it is possible to add terms non-linear in the gauge field  to the gauge-fixing term so as to modify the interactions as well. A convenient choice is
\be\label{H}
{\cal L}_{\rm g.f.} = -\frac{1}{2} {\rm Tr}(H)^2, \qquad H = \partial^\mu A_\mu + \frac{g}{\im\sqrt{2}} A^\mu A_\mu.
\ee
Indeed, by squaring (\ref{F-mat}) both the cubic and the quartic vertices contain two terms, coming from the symmetrisation of the $MN$ indices. The choice (\ref{H}) cancels one of the two terms in both cubic and the quartic vertex. Let us see this. Using the same trick as in (\ref{i-1}) we can rewrite the self-dual part of the curvature as
\be
F_{MN} = 2 \left( \partial_{MM'} A_N{}^{M'} + \frac{g}{\im\sqrt{2}}A_{MM'}A_N{}^{M'}\right) + \epsilon_{MN} H.
\ee
Then
\be
{\cal L}_{\rm YM} = - {\rm Tr}\left( \partial_{MM'} A_N{}^{M'} + \frac{g}{\im\sqrt{2}}A_{MM'}A_N{}^{M'}\right)^2 +\frac{1}{2} {\rm Tr}(H)^2
\ee
and hence
\be
{\cal L}_{\rm YM} + {\cal L}_{\rm g.f.} = -\frac{1}{2} {\rm Tr} (\partial^\mu A^\nu \partial_\mu A_\nu) + {\cal L}_{\rm YM}^{(3)}+ {\cal L}_{\rm YM}^{(4)}
\ee
with
\be\label{col-ym}
{\cal L}_{\rm YM}^{(3)} = \im g\sqrt{2} \,{\rm Tr} (\partial^{M}{}_{M'} A^{NM'} A_{MN'}A_N{}^{N'}),
\qquad
{\cal L}_{\rm YM}^{(4)} =\frac{g^2}{2} {\rm Tr}( A^{M}{}_{M'}A^{NM'}A_{MN'}A_N{}^{N'}).
\ee

We will also need the new vertex from (\ref{L2}) in the same matrix notation. Writing in terms of  gauge fields
defined  with one factor of coupling constant absorbed, we have
\be\label{col-new}
{\cal L}_{\rm new} = -\im \frac{\alpha g}{3  \sqrt{2}M^2} {\rm Tr}( F_A{}^B F_B{}^C F_C{}^A).
\ee

\subsection{Possibilities at higher order}

This representation of the new interaction also suggests a way to encode and classify the possibilities at higher powers of the curvature. Indeed, if $(T^a)_i{}^j$ is an $N\times N$ matrix, it is clear that in (\ref{col-new}) the curvature $F_A{}^B$ is thought of as a $2N\times 2N$ matrix
\be
F^a_A{}^B T^a_i{}^j := \hat{F}_{iA}{}^{jB}, 
\ee
so that the interaction (\ref{col-new}) is just a multiple of the trace of the cube of $\hat{F}$. Thus, at the order quartic in curvature we can have a single trace interaction ${\rm Tr}(\hat{F}^4)$, as well as the double trace interaction ${\rm Tr}(\hat{F}^2){\rm Tr}(\hat{F}^2)$. This way of classifying the new interactions can be clearly extended to higher orders. 

\subsection{Colour-ordered Feynman rules}

As we just saw, at quartic order and higher there are also multi-trace interactions in our class of theories. Because of this, even at tree level the scattering amplitudes for our theories no longer follow the simple single-trace pattern. This is not a cause of concern for us here, because this more involved structure will not show up in the low point amplitudes we compute. But one should keep in mind that the colour ordered rules for our theories are more involved than in YM because of the presence of multi-trace interactions. 

At the cubic interaction level all interactions are single-trace, and we can easily deduce the colour-ordered Feynman rules from (\ref{col-ym}) and (\ref{col-new}). We will be determining the tree-level scattering amplitudes using the recursion relations, and so we will only need the Feynman rules for these cubic vertices. We the colour stripped off, the YM cubic vertex is again represented as a sum of three  terms, with the derivative in each term acting on a different leg. We again use placeholders, but these are now colour stripped. We have for the YM cubic vertex
\be\label{3-ym}
-\im g\sqrt{2}\, k^M_{M'} A^{NM'}(k) A_{MN'}(p) A_N{}^{N'}(q) + {\rm 2\,\,terms},
\ee
and for the new vertex
\be\label{3-new}
-\im \frac{\alpha g}{\sqrt{2} M^2} (kA(k))_A{}^B (pA(p))_B{}^C (qA(q))_C{}^A,
\ee
where $2k^{M}{}_{M'} A^{NM'}(k)=(kA(k))^{MN}$. As before, the propagator is a factor of $1/\im k^2$ times a set of Kronecker delta's as well as spinor $\epsilon$'s. All momenta are assumed incoming. 

Note that the colour-ordered Feynman rules (\ref{3-ym}) and (\ref{3-new}) also admit a representation analogous to (\ref{v3-YM}) and (\ref{v3new}). In other words, one can still think in terms of the derivatives acting on legs of a diagram in terms of dots. One still has the key property (\ref{ident}) that two dots on the same internal line collapse the line. This will be important in what follows to estimate the high energy behaviour of the diagrams. 

\subsection{Colour-ordered cubic vertices on shell}

On-shell cubic vertices in any theory of massless particles vanish because of the special kinematics for three real null momenta satisfying the momentum conservation. However, as has become the common practice in the literature on scattering amplitudes, it is convenient to analytically continue the momenta to complex values in a certain way. One then obtains non-zero answers for the on-shell 3-point amplitudes. These then play a crucial role in the proof of constructibility, as they serve as seeds for the recursion that is used to construct all higher point amplitudes. 

Thus, we now evaluate the vertices (\ref{3-ym}) and (\ref{3-new}) on shell (with momenta continued to complex values), as this is the input of the recursion relation calculations. We first note that 
\be\label{kA}
(k\epsilon^-)^{MN} = 0, \qquad (k \epsilon^+)^{MN} = -2\im \, k^M k^N.
\ee
Hence the new vertex is only non-zero for the all plus configuration
\be\label{3-plus}
{\cal A}^{+++}_{\rm new} = \frac{4 \alpha g \sqrt{2}}{M^2} \ket{1}{2}\ket{2}{3}\ket{3}{1}.
\ee
Yang-Mills vertex, on the other hand, is zero for both the all plus and all-minus configurations. For the all minus configuration this is clear, as there must be at least one plus helicity so that the derivative term does not give zero. For the all plus configuration, there is a contraction of the primed indices of the two gauge field placeholders that are without the derivative, and this is what makes the amplitude zero. 

Let us compute YM vertex for the single plus two minus configuration. This is easy, because the plus helicity state should be inserted where the derivative acts. We get
\be\label{+--}
{\cal A}^{+--}_{\rm YM} = g\sqrt{2} \frac{\ket{1}{q}^2}{\ket{2}{q}\ket{3}{q}} \bra{2}{3} = g\sqrt{2} \frac{\bra{2}{3}^3}{\bra{1}{2}\bra{3}{1}},
\ee
where we have used the momentum conservation in the last step.

The opposite helicity configuration is a bit more complicated because there are two terms. We get
\be
{\cal A}^{-++}_{\rm YM} = -g\sqrt{2} \ket{2}{3}\left( \frac{\bra{1}{q}}{\bra{2}{q}} \frac{\ket{3}{q}}{\ket{1}{q}} +  \frac{\bra{1}{q}}{\bra{3}{q}} \frac{\ket{2}{q}}{\ket{1}{q}}\right) = -g\sqrt{2} \frac{\ket{2}{3}\bra{1}{q}}{\ket{1}{q}\bra{2}{q}\bra{3}{q}} ( \ket{3}{q}\bra{3}{q}+ \ket{2}{q}\bra{2}{q}).
\ee
We now use the momentum conservation to get
\be
{\cal A}^{-++}_{\rm YM} = g\sqrt{2} \frac{\ket{2}{3}^3}{\ket{1}{2}\ket{3}{1}},
\ee
which is just the complex conjugate of (\ref{+--}) as it should be. 

\section{Recursion relations}
\label{sec:Recur}

In this section we explain how the tree level scattering amplitudes can be computed using recursion relations. First, the maximally helicity violating (MHV) amplitudes, now defined as those with just two plus helicities and the rest minus are unchanged from the YM case. Then the amplitudes with at least one minus can be computed using the usual BCFW recursion where a pair of momenta are shifted. The all plus amplitudes, which are now non-zero, are computed using the Risager's shift BCFW recursion. 

\subsection{Amplitudes with all negative helicity gluons}

It is easy to see that amplitudes for all negative helicity gluons, or just a single positive helicity configuration, continue to be zero for our theories as they are in the case of YM. Indeed, as usual, we choose all reference spinors of negative helicity gluons to be the same. In this case, the only hope to contract these helicity spinors and get a non-zero result is to use the momenta in the numerators of the diagrams, coming from the derivatives present in the vertices. This is best achieved by having as many derivatives as possible in a diagram. And the maximum number of derivatives in a diagram is achieved by the 3-valent graphs. 

It first looks like the new vertex (\ref{3-new}) should change the situation drastically, because it contains three derivatives instead of single derivative in the YM vertex case. But because of the identity (\ref{ident}) there cannot be too many copies of this vertex in a diagram. Indeed, the identity (\ref{ident}) implies that there can be at most a single dot on every internal line of a diagram. Consider now a 3-valent diagram for $n$ negative helicity gravitons. There are $n-3$ internal lines. Note that the dot cannot appear on the external lines, because it will then give zero applied to a negative helicity graviton. So, the maximal total number of derivatives in a diagram for all negative helicity gluons is $n-3$, which is not enough to absorb all $n$ reference spinors. This shows that all negative helicity amplitudes are zero.

A similar argument is at work for the configuration with at most single positive helicity gluon. We can then choose the reference spinors of negative helicity gluons to be the unprimed momentum spinor of the positive helicity gluon. Again, we have $n$ unprimed spinors to contract, while avoiding contracting them to each other. Now that we have one positive helicity gluon, we can have one derivative on an external leg, raising the total number of possible derivatives in a 3-valent diagram to $n-2$. This is still not enough to contract all $n$ reference spinors, hence giving a zero result for such amplitudes. Below we will also treat this case using the BCFW recursion. 

\subsection{Amplitudes with at least one negative and one positive helicity gluon}

Let us now consider more general amplitudes. Let us assume that there is at least one negative helicity gluon, and at least one positive. The case not covered is thus just that of all positive helicity configuration. This will be treated separately. 

We would now like to analyse the behaviour of amplitudes under the BCFW shift, where the shift is applied to a negative-positive pair of gluons. Let us label the gluons so that a chosen negative helicity gluon has label $1$, and the chosen positive gluon label $n$. We then perform the BCFW shift \cite{Britto:2005fq}
\be\label{shift}
1\to \hat{1}=1+ z\,n, \qquad n'\to \hat{n}' = n'- z1'.
\ee
Under this shift the helicity spinors (\ref{helicity}) of both gluons change and behave as $1/z$ for large complex parameters $z$. 

We can now count the powers of $z$ appearing in a generic Feynman diagram shifted as in (\ref{shift}). We are interested in the large $z$ behaviour, and would like to show that the amplitudes go as $1/z$ in this limit. Thus, as in the previous subsection, we need to be concerned about the derivatives present in the vertices. The worst case scenario is when there are only cubic vertices present. However, as we already used in the previous section, in this case there can at most be $n-3$ derivatives on the internal lines. This equals to the number of propagators. Each propagator on a curve connecting vertices $1$ and $n$ will contribute a factor of $1/z$, and the derivatives on these internal lines will each contribute a factor of $z$. Thus, there will be a cancellation of factors of $z$ from the internal lines. The last thing to note is that both external lines connecting to $1$ and $n$ cannot contain a derivative. Indeed, as one of the helicities is negative there cannot be a derivative on that external line. Thus, we see that the helicity spinors each give a $1/z$ factor, while there is at most a single derivative on the external line, giving another factor of $z$. The above discussion can be summarised in a formula as follows. Assuming the number of internal lines connecting the shifted pair $(1^{-},n^{+})$ is $I$, we get the following $z$-behaviour:
\begin{equation}
\left(\frac{1}{z}\right)^{I}z^{I}\times z\times\left(\frac{1}{z}\right)^{2}=\frac{1}{z}.
\end{equation}
Overall, we see the desired $1/z$ behaviour of such amplitudes. An illustration of this shifting is provided by Fig 1.
\begin{figure}
\begin{center}
\includegraphics[width=2.5in]{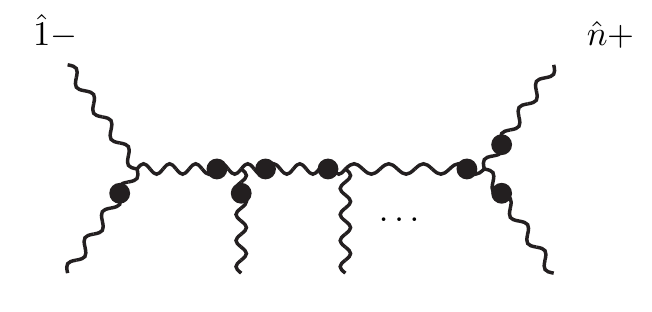}
\caption{Standard BCFW shifting on the pair $(1^{-},n^{+})$}
\end{center}
\label{fig:generic-pair-shifting}
\end{figure}

It is also not hard to deduce the desired $1/z$ behaviour of the amplitudes from an argument analogous to the one used in \cite{ArkaniHamed:2008yf}. This is the background field method argument, where the action is expanded to second order in the ``hard'' field, on a background of ``soft'' particles. To obtain the $1/z$ behaviour one should just note that the interaction of the hard perturbation with the soft background contains at most a single derivative of the hard field. This is trivial to see in the YM case. In our ``deformed'' case, there is also a second-derivative interaction given by
\be
\frac{\partial^2 f}{\partial F^a_{MN} \partial F^b_{PQ}} (\partial A)^a_{MN} (\partial A)^b_{PQ},
\ee
where the derivative of the function $f$ is evaluated at the background. The other, subleading interaction terms contains at most a single derivative. It seems that we now have a dangerous second-derivative interaction of the hard and soft fields. What saves the game is that for the helicity configuration $-+$ one of the hard fields is necessarily a negative helicity field, for which the object $(\partial A)^a_{MN}$ vanishes. So, for the $-+$ hard field configuration, the interaction with the soft field starts with a first derivative term, which is what is needed to get the desired $1/z$ behaviour. 

The conclusion is then that the amplitudes with at least one minus and one plus gluon behave as $1/z$ under the BCFW shift (\ref{shift}). Because of this they can be determined by the BCFW recursion relation, with only on-shell vertices needed to compute the amplitudes. However, to run the BCFW recursion, generically we also need to know the all plus amplitudes. 

\subsection{MHV amplitudes}

We define the MHV amplitudes as those with exactly two positive helicity gluons, and an arbitrary number of the negative helicity particles. These are the simplest non-vanishing amplitudes. To determine these amplitudes we do not need to know the all plus amplitudes, and so we can discuss them already now. 

Let us first note that we can use the BCFW recursion to show that amplitudes with just one plus helicity gluon continue to be zero. Indeed, such amplitudes are cut by BCFW into two copies of lower point amplitudes of the same kind
\begin{equation}\nonumber
\cA(1^{-},2^{-},\dots,n^{+})=\sum_{k}\cA(\hat{1}^{-},2^{-},\dots,k^{-},\hat{P}^{+})\frac{1}{P^{2}}\cA(-\hat{P}^{-},k+1^{-},\dots,\hat{n}^{+})=0,
\end{equation}
and thus are zero by induction.

It is now easy to see that the MHV amplitudes continue to be given by the same expressions as arise in Yang-Mills theory. This follows from the applicability of the BCFW recursion relations in this case, as there is clearly at least one minus and at least one plus helicity gluon. Given that the all minus and all-except-one minus amplitudes are all zero, the on-shell relation for MHV amplitudes prescribes 
\begin{eqnarray}\nonumber
\cA(1^{-},2^{-},3^{-}\dots,i^{+},\dots,n-1^{-},n^{+}) & = & \cA(\hat{1}^{-},2^{-},\dots,i^{+},\dots,\hat{P}^{+})\frac{1}{P^{2}}\cA(-\hat{P}^{-},n-1^{-},\hat{n}^{+})\\
 &  & +\cA(1^{-},2^{-},\hat{P}^{+})\frac{1}{P^{2}}\cA(-\hat{P}^{-},3^{-},\dots,i^{+},\dots,\hat{n}^{+}).\nonumber
\end{eqnarray}
The first line here vanishes at the pole, while the second line gives the same contribution as in the original Yang-Mills theory. So, one is never using any new amplitudes in this recursion, and the arising MHV amplitudes are just those of Yang-Mills theory.  

\subsection{All plus amplitudes}

Our above arguments based on the BCFW shift (\ref{shift}) are not applicable to the all plus helicity configurations. And indeed, we should not expect these amplitudes to decay under the BCFW shifts. Consider for example the (\ref{shift}), but now applied to two positive helicity gluons. It is clear that under this shift the basic 3-point amplitude (\ref{3-plus}) behaves as $O(z)$ for large $z$. Thus, the simple BCFW shift acting on just a pair of gluons is certainly not going to work for the all plus amplitudes. 

However, there is another possibility, first used in work by Risager \cite{Risager:2005vk}, where all gluons are shifted. Thus, let us shift the primed momentum spinors of all the positive helicity particles as
\be\label{riss}
i'\to \hat{i}' = i' + z f_i q', \qquad \sum_i i f_i = 0.
\ee
In other words, we shift each primed momentum spinor in the direction of the reference spinor used in the helicity spinors. This is done in such a way that the momentum conservation is respected. An explicit form for the coefficients $f_i$ will be given below at four points. 

We will see that the shift works for our purposes, because it can be used to determine the parts of the $n$-all plus amplitude that arise from lower order amplitudes. However, generically at $n$-points there will also be parts of the amplitude that cannot be determined by the recursion, the so-called BCFW boundary terms. However the sources of boundary terms can be readily identified. Indeed, assuming for simplicity that we deal with the theory with just a single new interaction (\ref{L2}), it is easy to see that the only diagrams not exhibiting the $1/z$ behaviour are those constructed solely from new cubic vertices. The contribution to BCFW boundary is then simply given by the sum of diagrams that completely collapse
\be\nonumber
\includegraphics[width=5in]{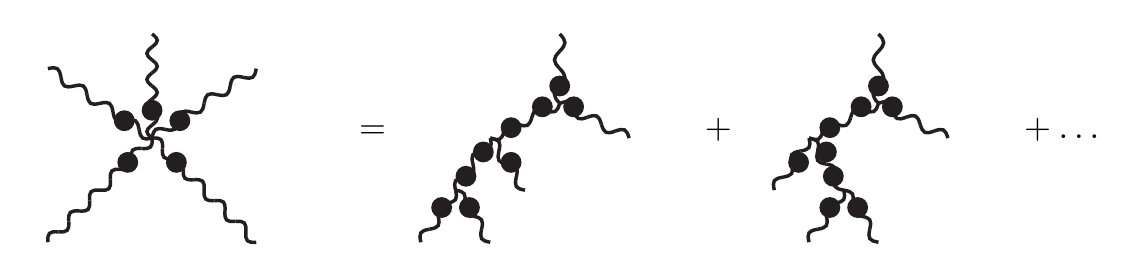}
\ee
This sum can be evaluated explicitly, as the terms that appear are just spinor contractions $\ket{i}{j}$. An example at four points  will be considered below. 

Alternatively, one can note that the boundary terms that appear are just multiples of the higher-valent vertices that anyway have to be included into the theory. So, one learns that the parts of the $n$-point all plus amplitude that are not determined by the BCFW are just the  $n$-valent vertices evaluated on shell. These parts come with new coupling constants and should simply be added to the those parts that are determined by the BCFW recursion with Risager's shift to obtain the full amplitude. 

To see that the rest of the diagrams indeed converge under the all leg shift (\ref{riss}), let us apply the collapsing rule until there are no internal lines to collapse. If no internal lines are left, we are in the situation of the boundary term just considered. So, we are left with diagrams with ``effective'' higher-valent vertices, and internal lines with just a single dot or no dots. The lines with a single dot are dangerous, because there is a factor of $1/z$ coming from the denominator, and a possible factor of $z$ from the derivative. Lines without dots are not dangerous, as they immediately give a $1/z$ behaviour to the diagram. Thus, we can concentrate on dangerous diagrams where all internal lines carry a single dot. 

To see the $z$-behaviour of these diagrams, let us consider a line with a single dot on it. This dot is on one end of the line, and so there is no dot on the other end. Thus, the end of the line that does not have a dot is contracted into some effective vertex of the type 
\be\nonumber
\includegraphics[width=1.5in]{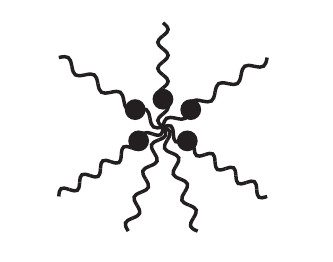}
\ee
Note that there is necessarily two (or more generally even number of) legs without dots, as otherwise the primed index of the placeholder connections will have nowhere to contract. We now follow the other leg of this vertex where there is no dot. If there is more than a single pair of undotted legs, we follow the leg to which the primed index of the placeholder for the line we came from is contracted to. We concluded that the dangerous diagrams are those where every internal line has one dot, so there is a dot on the other end of this second line. 

What we now have is a pair of internal lines, each with a single dot on them, connecting at an effective vertex. What is important for us is that in this effective vertex the primed indices of the two gauge field placeholders contract. This means that the numerator arising from the dot on the first line is contracted to the numerator on the second line in their primed indices. As the result, the would be dangerous $z^2$ behaviour of the product does not arise, because it is necessarily proportional to $\bra{q}{q}=0$. This establishes that there are no other boundary terms, and everything apart from the fully collapsed diagrams falls off as $1/z$ or even faster as $z\to\infty$. So, we can use Risager's shift to determine the ``complicated'' parts of the $n$-point all plus amplitudes. 

\section{4-point examples}
\label{sec:4-point}

The purpose of this section is to illustrate the general statements of the previous section on the examples of 4-point amplitudes. Some amplitudes with $F_{\rm SD}^3$ interaction have been previously computed in the literature; see in particular \cite{Broedel:2012rc}. So, there is no claim for novelty in this section, and it is given more for illustration on the types of amplitudes that arise. 

\subsection{$(--++)$ MHV amplitude}

As we have already discussed, there are no all minus and $---+$ amplitudes, so the first one to consider is the MHV amplitude $(--++)$. We expect that it is given by the same expression as in the usual YM theory. 

From the Feynman diagram calculation perspective it may seem that the following two new diagrams may contribute to this amplitude
\be\nonumber
\includegraphics[width=4in]{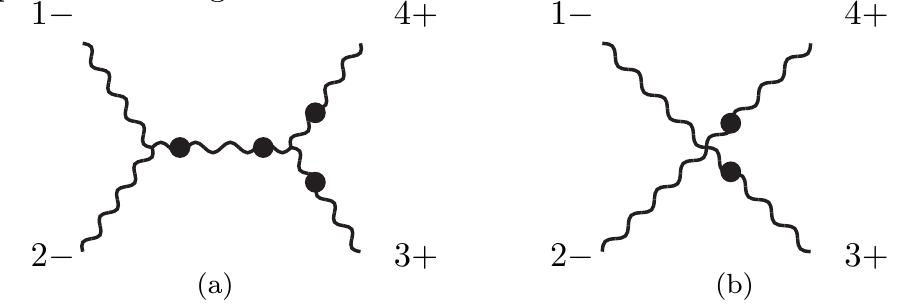}
\ee
However, it can be checked that they precisely cancel each other in view of the collapsing rule. So, the MHV amplitude is completely determined by the original Yang-Mills diagrams. 

The same conclusion can also be derived from the on-shell relation perspective. As we discussed in the previous section, we can apply the $-+$ shift (\ref{shift}). The only possible contribution is then 
\begin{equation}
\cA(1^{-},2^{-},3^{+},4^{+})=\cA(\hat{1}^{-},2^{-},\hat{P}^{+})\frac{-i}{P^{2}}\cA(-\hat{P}^{-},3^{+},\hat{4}^{+}),
\end{equation}
with $P^2=2\ket{1}{2}\bra{1}{2}$. Thus, the on-shell recursion relation of the MHV amplitude does not contain the new $3$-point all plus amplitude and is therefore unchanged in the deformed theory.
\begin{equation}
\cA(1^{-},2^{-},3^{+},4^{+})=(-i) g^{2}\frac{\bigl[12\bigr]^{4}}{\bigl[12\bigr]\bigl[23\bigr]\bigl[34\bigr]\bigl[41\bigr]}=(-i) g^{2}\frac{\bigl\langle34\bigr\rangle^{4}}{\bigl\langle12\bigr\rangle\bigl\langle23\bigr\rangle\bigl\langle34\bigr\rangle\bigl\langle41\bigr\rangle},
\end{equation}
 and similarly for the non-adjacent MHV amplitude.

\subsection{$(-+++)$ amplitude}

The first $4$-point amplitude that is affected by the deformation
is $\cA(1^{-},2^{+},3^{+},4^{+})$. At first sight we have one diagram
that may contribute to the BCFW boundary
\be\nonumber
\includegraphics[width=1.6in]{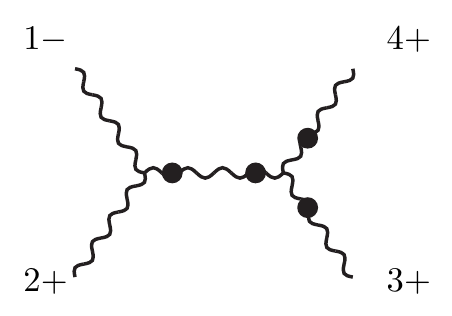}
\ee
However as in the previous discussion this diagram behaves as $1/z$ because of the dots collapsing rule. Therefore, the amplitude is given by an on-shell relation as
\begin{equation}
\cA(1^{-},2^{+},3^{+},4^{+})=\cA(\hat{1}^{-},2^{+},\hat{P}^{+})\frac{-i}{P^{2}}\cA(-\hat{P}^{-},3^{+},\hat{4}^{+})+\cA(\hat{1}^{-},2^{+},\hat{P}^{-})\frac{-i}{P^{2}}\cA(-\hat{P}^{+},3^{+},\hat{4}^{+})
\end{equation}
The first term is not contributing because $\cA(\hat{1}^{-},2^{+},\hat{P}^{+})$ vanishes at the pole, whereas the second term gives
\begin{equation}
\cA(\hat{1}^{-},2^{+},\hat{P}^{-})\frac{-i}{P^{2}}\cA(-\hat{P}^{+},3^{+},\hat{4}^{+})= \sqrt{2} g\frac{\bigl[\hat{P}1\bigr]^{3}}{\bigl[12\bigr]\bigl[2\hat{P}\bigr]}\,\frac{-i}{P^{2}}\,2^{2} \sqrt{2} \frac{\alpha g}{M^{2}} \bigl\langle\hat{P}3\bigr\rangle\bigl\langle34\bigr\rangle\bigl\langle4\hat{P}\bigr\rangle
\end{equation}
Note that this is different from the usual Yang-Mills theory where we do not have a non-vanishing all plus $3$-point sub-amplitude. Substituting the on-shell value for internal momentum $\hat{P}=k_2(\ket{4}{1}/\ket{4}{2} k_1'+k_2')$ we get
\begin{equation}\label{ym-mppp}
\cA(1^{-},2^{+},3^{+},4^{+})= i  \frac{4\alpha g^{2} }{M^{2}} \frac{\bigl\langle23\bigr\rangle\bigl\langle34\bigr\rangle}{\bigl\langle12\bigr\rangle\bigl\langle14\bigr\rangle}\bigl\langle24\bigr\rangle^{2}.
\end{equation}

\subsection{The all plus amplitude}

When all four external legs carry plus helicities, the all plus amplitude diverges under the standard BCFW shift (\ref{shift}). Let us instead consider the following generalisation of the Risager
shift on all four external legs
\begin{equation}
\begin{array}[t]{cccccc}
\hat{k}_{1}= & k_{1}k_{1}' & +0 & +zk_{1}\bigl\langle34\bigr\rangle \eta' & +z\,k_{1}\bigl\langle24\bigr\rangle\eta' & +z\,k_{1}\bigl\langle23\bigr\rangle \eta'\\
\hat{k}_{2}= & k_{2}k_{2}' & +zk_{2}\bigl\langle34\bigr\rangle \eta' & +0 & +zk_{2}\bigl\langle41\bigr\rangle\eta' & +zk_{2}\bigl\langle31\bigr\rangle\eta'\\
\hat{k}_{3}= & k_{3}k_{3}' & +zk_{3}\bigl\langle42\bigr\rangle\eta' & +zk_{3}\bigl\langle41\bigr\rangle\eta' & +0 & zk_{3}\bigl\langle12\bigr\rangle\eta'\\
\hat{k}_{4}= & k_{4}k_{4}' & +zk_{4}\bigl\langle23\bigr\rangle\eta' & +zk_{4}\bigl\langle13\bigr\rangle\eta' & +zk_{4}\bigl\langle12\bigr\rangle\eta' & +0
\end{array}\label{eq:all-leg-shifting}
\end{equation}
Let us also choose all reference spinors to be identical to the same $\eta'$ introduced to define the shiftings. Since only the anti-holomorphic (primed) spinor are shifted, the polarisations are unchanged. 
 
Then the diagrams consisting of two new cubic vertices is finite for large $z$ and therefore contributes as the BCFW boundary term(s), see below. All other diagrams go to zero for large $z$, and can be calculated separately. Indeed, the $s$-channel residue is given by the following two $\overline{\text{MHV}}$/$(+++)$ combinations, which both converge under Risager shift (but diverge in the standard BCFW setting)
\begin{eqnarray}
A(\hat{1}^{+},2^{+},\hat{P}^{-})\frac{-i}{P^{2}}A(-\hat{P}^{+},3^{+},\hat{4}^{+}) & = & \sqrt{2}g\frac{\bigl\langle12\bigr\rangle^{3}}{\bigl\langle2\hat{P}\bigr\rangle\bigl\langle\hat{P}1\bigr\rangle}\,\frac{-i}{P^{2}}\, 2^{2}\sqrt{2} \frac{\alpha g}{M^{2}} \bigl\langle\hat{P}3\bigr\rangle\bigl\langle34\bigr\rangle\bigl\langle4\hat{P}\bigr\rangle\\
A(\hat{1}^{+},2^{+},\hat{P}^{+})\frac{-i}{P^{2}}A(-\hat{P}^{-},3^{+},\hat{4}^{+}) & = & 2^{2}\sqrt{2} \frac{\alpha g }{M^2} \bigl\langle\hat{P}1\bigr\rangle\bigl\langle12\bigr\rangle\bigl\langle2\hat{P}\bigr\rangle\,\frac{-i}{P^{2}}\,\sqrt{2} g \frac{\bigl\langle34\bigr\rangle^{3}}{\bigl\langle4\hat{P}\bigr\rangle\bigl\langle\hat{P}3\bigr\rangle}.\nonumber
\end{eqnarray}
For brevity we denote the shiftings (\ref{eq:all-leg-shifting}) taken on legs $1$ and $2$ as
\begin{eqnarray}
\hat{k}_{1}^{'} & = & k_{1}'+z\, c_{1}\eta,\\
\hat{k}_{2}^{'} & = & k_{2}'+z\, c_{2}\eta'\nonumber 
\end{eqnarray}
The on-shell condition at the $s$-channel pole allows us to write 
\begin{equation}
\hat{P}=\frac{1}{c_{1}\bigl[2\eta\bigr]-c_{2}\bigl[1\eta\bigr]}\left(k_{1}\bigl[1\eta\bigr]+k_{2}\bigl[2\eta\bigr]\right)\left(c_{1}k_{2}'-c_{2}k_{1}'\right)
\end{equation}
The $s$-channel residue can then be readily obtained by substituting $\hat{P}=k_{1}\bigl[1\eta\bigr]+k_{2}\bigl[2\eta\bigr]=-k_{3}\bigl[3\eta\bigr]-k_{4}\bigl[4\eta\bigr]$. We get
\be
(-\im) \frac{4\alpha g^2}{M^2} \frac{\ket{1}{2}}{\bra{3}{4}\bra{1}{\eta}\bra{2}{\eta}\bra{3}{\eta}\bra{4}{\eta}} \left( c_{12}^2 + c_{34}^2 \right),
\ee
where we introduced the following anti-symmetric quantities:
\be
c_{ij} := \bra{\eta}{i}\ket{i}{j}\bra{j}{\eta}.
\ee
Repeating the same calculation for the $t$-channel, we find the total amplitude   
\be\nonumber
A_{singular}(1^{+},2^{+},3^{+},4^{+})= (-i) \frac{4\alpha g^{2}}{M^{2}} \frac{1}{\bra{1}{\eta}\bra{2}{\eta}\bra{3}{\eta}\bra{4}{\eta}} \left(\frac{\ket{1}{2}}{\bra{3}{4}} \left( c_{12}^2 + c_{34}^2 \right)+\frac{\ket{1}{4}}{\bra{3}{2}} \left( c_{14}^2 + c_{32}^2 \right)\right).
\ee
Using the momentum conservation this becomes
\be\label{pppp-1}
A_{singular}(1^{+},2^{+},3^{+},4^{+})=(-\im) \frac{4\alpha g^2}{M^2} \frac{\ket{1}{2}}{\bra{3}{4}\bra{1}{\eta}\bra{2}{\eta}\bra{3}{\eta}\bra{4}{\eta}} \left( c_{12}^2 + c_{34}^2 - c_{14}^2-c_{23}^2\right).
\ee
The momentum conservation implies that we have the following identities satisfied by the $c$'s
\be
\sum_j c_{ij} = 0.
\ee
Because of this we can rewrite the quantity in brackets in (\ref{pppp-1}) as
\be
(c_{12}-c_{14})(c_{12}+c_{14}) + (c_{34}-c_{23})(c_{34}+c_{23}) = (c_{12}-c_{14})(-c_{13})+(-c_{31})(c_{34}+c_{23}) = \\ \nonumber
c_{13}(c_{34}+c_{23}-c_{12}+c_{14})=2 c_{13} c_{42} = 2\bigl[1\eta\bigr]\bigl[2\eta\bigr]\bigl[3\eta\bigr]\bigl[4\eta\bigr]\,\bigl\langle13\bigr\rangle\bigl\langle42\bigr\rangle.
\ee
Thus, the singular part of the $4$-point all plus amplitude is finally given by
\begin{equation}
A_{singular}(1^{+},2^{+},3^{+},4^{+})= (-i) \frac{8 \alpha g^{2}}{M^{2}} \frac{1}{\bigl[34\bigr]}\bigl\langle13\bigr\rangle\bigl\langle42\bigr\rangle\bigl\langle12\bigr\rangle.
\end{equation}

The BCFW boundary term, on the other hand, can be derived from a direct calculation of the $s$ and $t$-channel diagrams consisting of new cubic vertices only. In viewing of the dot collapsing rule, we can denote the sum of these two diagrams as
\be\nonumber
\includegraphics[width=4in]{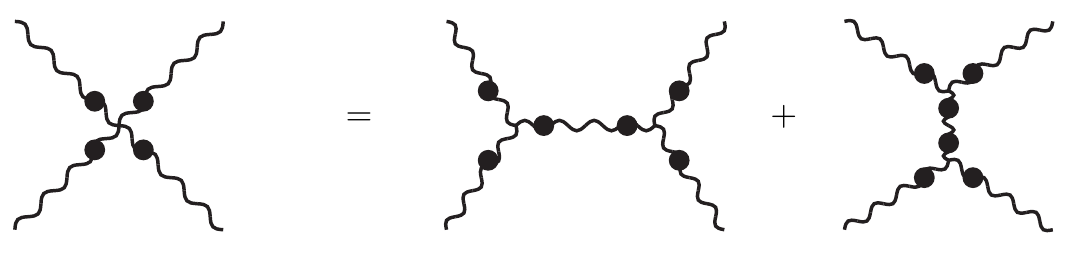}
\ee
The boundary term is given by contracting these diagrams with the positive helicity polarisations. For the first one we get
\be\nonumber
\im \frac{16\alpha^2 g^2}{M^4} \ket{1}{2}\ket{3}{4}\left( \ket{1}{3}\ket{2}{4}+  \ket{2}{3}\ket{1}{4}\right),
\ee
and the second one gives
\be\nonumber
\im \frac{16\alpha^2 g^2}{M^4} \ket{1}{4}\ket{3}{2}\left( \ket{1}{3}\ket{4}{2}+  \ket{4}{3}\ket{1}{2}\right).
\ee
The sum is
\begin{equation}\label{4-p}
A_{boundary}(1^{+},2^{+},3^{+},4^{+})= \im\,\frac{16\alpha^{2}g^{2}}{M^{4}} \Big[ \ket{1}{3}^2\ket{2}{4}^2-2\ket{1}{2}\ket{2}{3}\ket{3}{4}\ket{4}{1}\Big].
 \end{equation}
 It is clear that the boundary term in the all plus amplitude is of the form that can be obtained by evaluating the 4-vertex of type $F^4$ on shell. As we discussed previously, at the quartic level there are two possible interactions, one single-trace, one double-trace. The last term in (\ref{4-p}) is precisely of the type that would be obtained by evaluating the single-trace quartic interaction on-shell. The first term is of the double-trace type, but here one must be careful to include all colour orderings to interpret the arising colour structure correctly. In any case, both contributions to the boundary term are of the type arising by evaluating the quartic vertex on-shell.
 
One now has two options. One option is simply to interpret this boundary contribution as a shift of of the new couplings that get added to the theory at the 4-point level. Then one simply forgets about this boundary contribution, and adds a set of all plus amplitudes with arbitrary couplings. 
 
 Another option, possibly more interesting, could be to require that the amplitudes are completely determinable by the recursion, with no new couplings to be added apart from the already existing $\alpha$ coupling. Thus, one can require that the new interactions of type $F^4$ that are added to the Lagrangian are designed so that there is only a contribution to the 4-plus amplitude behaving as $1/z$ under the all momentum shift. In other words, one can decide to cancel the above boundary term with what comes from the interactions of type $F^4$. If one continues to add new interactions at every order so as to have the $1/z$ falloff of the all plus amplitudes, one obtains a set of amplitudes described just by two couplings, the YM coupling $g$ and the new coupling $\alpha$. One can also decide to add a finite set of higher-order couplings, and then request the $1/z$ behaviour of all plus amplitudes starting from some order $n$. This way one will obtain amplitudes parameterised by a finite number of couplings. All these options are possible due to the constructible character of amplitudes for our theories. 

\section{The double copy structure}
\label{sec:double}

The purpose of this section is to point out that the 4-gluon amplitudes obtained in the previous section continue to be related to the 4-graviton amplitudes for deformed GR known from \cite{Delfino:2012aj} via the double copy pattern. The fact that non-renormalisable $F^3$ gauge theory interactions continue to square to gravity amplitudes is not new, and it has been discussed in e.g. \cite{Cohen:2010mi} and \cite{Broedel:2012rc}. The latter reference also contains a discussion of the colour-kinematics duality, which we do not go into. What is new in the present paper is the purely chiral context where both the gauge theory and gravity sides are augmented by only the chiral half of the amplitudes. 

\subsection{Graviton scattering amplitudes for deformations of GR}

In this paper we will only look at the relation between 4-particle amplitudes, leaving the general case to future study. 

The 4-graviton scattering amplitudes in a family of ``deformations'' of General Relativity studied in \cite{Delfino:2012aj} are as follows. The all minus and one plus amplitudes continue to be zero as in GR. The $--++$ MHV amplitude is unchanged from what it is in GR, and is given by
\be
{\cal M}^{--++} = \im \frac{1}{M_p^2} \ket{3}{4}^6 \frac{1}{\ket{1}{3}\ket{1}{4}\ket{2}{3}\ket{2}{4}} \frac{\bra{1}{2}}{\ket{1}{2}},
\ee
with the Planck mass being $M_p^2 = 1/16\pi G$. The $-+++$ amplitude is non-zero in deformed GR, and is given by
\be\label{-+++}
{\cal M}^{-+++} = - \im \frac{27\beta^2 }{4M_p^6} \bra{1}{2} \frac{\ket{2}{3}^3 \ket{2}{4}^3\ket{3}{4}^2 }{\ket{1}{2}\ket{1}{3}\ket{1}{4}},
\ee 
where $\beta$ is the  parameter for a particular one-parameter family of deformations described in the Appendix in \cite{Delfino:2012aj}. Finally, the all plus amplitude consists of two contributions
\be\label{++++}
{\cal M}^{++++} = \im \frac{135\beta^2 }{2M_p^6}\ket{1}{2}\ket{1}{3}\ket{1}{4}\ket{2}{3}\ket{2}{4}\ket{3}{4} \frac{\ket{3}{4}}{\bra{1}{2}} \\ \nonumber
+ \im \frac{9\beta^3}{2M_p^8} \left( \ket{1}{2}^2\ket{2}{3}^2\ket{3}{4}^2\ket{4}{1}^2+\ket{1}{2}^2\ket{2}{4}^2\ket{4}{3}^2\ket{3}{1}^2+\ket{1}{3}^2\ket{3}{2}^2\ket{2}{4}^2\ket{4}{1}^2\right).
\ee
The first of these contributions comes from a Feynman diagram calculation and has $s,t,u$ plane singularities. The second contribution is just obtained by evaluating the quartic vertex on-shell. The coupling constant appearing in front of the quadratic vertex does not have to be related to the cubic vertex couplings. Thus, the coefficient in front of the term in the second line is in general arbitrary.

\subsection{The double copy structure of the MHV amplitude}

Here we review the usual gravity as square of YM story. A good reference for this is \cite{Bern:2002kj}. At the level of the 4-point amplitudes we have
\be\label{double-copy}
M_p^2 {\cal M}^{--++}(1,2,3,4) = \im \frac{1}{g^2} {\cal A}^{--++}(1,2,3,4) s_{12}  \frac{1}{g^2} {\cal A}^{--++}(1,2,4,3),
\ee
where here $s_{12}=\ket{1}{2}\bra{1}{2}$. Given the above explicit form of the amplitudes, it is a simple exercise to verify this relation. We have a different sign in this formula as compared to one in \cite{Bern:2002kj}, which is due to our different convention for the helicity spinors. 

\subsection{The double copy structure of the new amplitudes}

Let us apply the double copy relation (\ref{double-copy}) to the $-+++$ amplitudes. In this case the left hand-side gives
\be\nonumber
-\im \left(\frac{4\alpha}{M^2} \frac{\ket{3}{2}\ket{3}{4}}{\ket{1}{2}\ket{1}{4}} \ket{2}{4}^2\right) \ket{1}{2}\bra{1}{2}  \left(\frac{4\alpha}{M^2} \frac{\ket{4}{2}\ket{4}{3}}{\ket{1}{2}\ket{1}{3}} \ket{2}{3}^2\right) = \im \frac{16\alpha^2}{M^4} \bra{1}{2} \frac{\ket{2}{3}^3\ket{2}{4}^3\ket{3}{4}^2}{\ket{1}{2}\ket{1}{3}\ket{1}{4}},
\ee
apart from an extra minus sign that is likely due to our different helicity conventions, which starts to play a role when there is a different number of positive and negative helicity particles, and this equals to the left-hand-side in (\ref{double-copy}) with the gravitational amplitude given by (\ref{-+++}) if we identify
\be\label{gr-YM}
 \frac{16\alpha^2}{M^4} = \frac{27\beta^2}{4 M_p^4}.
 \ee

Let us also work out the square of the (singular part of the) gauge theory $++++$ amplitude. The left-hand-side of (\ref{double-copy}) in this case gives
\be\nonumber
-\im \left(\frac{8\alpha}{M^2} \frac{\ket{1}{3}\ket{2}{4}\ket{3}{4}}{\bra{1}{2}} \right) \ket{1}{2}\bra{1}{2} \left(\frac{8\alpha}{M^2} \frac{\ket{1}{4}\ket{2}{3}\ket{4}{3}}{\bra{1}{2}} \right)= \im \frac{64\alpha^2}{M^4} \ket{1}{2}\ket{1}{3}\ket{1}{4}\ket{2}{3}\ket{2}{4}\ket{3}{4} \frac{\ket{3}{4}}{\bra{1}{2}} .
\ee
With the identification (\ref{gr-YM}), modulo the numerical coefficient, this matches the first term in the gravity amplitude (\ref{++++}). 

For the second term in (\ref{++++}) that is obtained by a simple evaluation of a 4-vertex on shell, one should not expect the pattern (\ref{double-copy}) to continue. Indeed, the reason for $s_{12}$ in (\ref{double-copy}) is to cancel an extra internal line propagator when combining two YM amplitudes. In the case when an amplitude is obtained directly from a 4-vertex, with no internal propagator to be cancelled, one should expect the gravity amplitude to be simply the square of the YM one, as it happens for example for the 3-point amplitudes. And indeed, we see that the gravity all plus amplitude is a sum of three terms, each being a square of a colour-ordered gauge theory amplitude. The sum is necessary to produce a completely permutation-invariant expression from the ordered gauge theory amplitudes. 

So, overall we see that the gravity as a square of gauge theory pattern is at play for all the amplitudes. There are however subtleties about numerical coefficients still to be understood. Presumably, to reproduce the gravity amplitudes together with the correct numerical coefficient one should subtract the unwanted modes arising by tensoring the two spin one states; see \cite{Broedel:2012rc} for a related discussion. It is also necessary to understand what replaces the usual double copy rules for the all plus amplitudes, with their two different types of terms. We leave all this to future work. 

\section{The first look at renormalisation}
\label{sec:renorm}

As we have already mentioned in the Introduction, in many respects our theories behave as the renormalisable Yang-Mills theory from which they originated. This refers to the property of our perturbation theory that guarantees that there is at most a single derivative on every internal Feynman diagram line. In this aection we would like to analyse consequences of this for the power-counting, as well as study divergences of some simple diagrams. We will see that there are divergences of a new type, not present in YM, as can be expected from a power-counting non-renormalisable theory. However, at least at one loop these divergences can be absorbed by a local field redefinition. 

\subsection{Choice of the renormalisation scheme}

As it was clear from our motivation section, and from the analysis of the tree-level scattering amplitudes, the ``collapsing'' property (\ref{ident}) plays a very useful role in simplifying the structure of the perturbation theory. Thus, we definitely want to keep this property operating also at the loop level. 

However, here a complication arises. The most convenient renormalisation scheme to use in the context of such a theory as Yang-Mills is the dimensional regularisation. This is however not directly applicable to theories with spinors in them, because of the difficulties in continuing the notion of a spinor to arbitrary (non-integer) dimension. To put this problem differently, our theories use self-dual objects, and this is clearly a 4-dimensional notion, not directly extendible to arbitrary dimension. 

We will circumvent these problems in the following way. First, we note that we only need to dimensionally continue the loop momenta in our diagram, as it is the divergent integrals over these loop momenta that need to be regularised. So, the only object from the Feynman rules spelled out above that needs to be defined in general dimension is the momentum $k_{AA'}$ on an internal line of a diagram. In four dimensions this object is defined as $k_{AA'}= - \theta^\mu_{AA'} k_\mu$, where $\theta^\mu_{AA'}$ are the soldering forms, $k_\mu$ is the usual 4-momentum vector, and the minus sign is due to our choice of conventions, which are those of section 6 of \cite{Delfino:2012aj}.

Now, in $d$ dimensions one can define the notion of momentum $k_\mu$ in the usual way. To define the notion of $k_{AA'}$ we therefore need to dimensionally continue the soldering forms $\theta^\mu_{AA'}$. There is a known way to do this, familiar from the literature on the dimensional regularisation of theories with fermions. Indeed, a similar problem arises e.g. in QED, where one needs to dimensionally continue the $\gamma$-matrices. One does this in a such a way as to preserve the Clifford algebra defining relation. With the soldering forms being just the $2\times 2$ building blocks of the $\gamma$-matrices, one can adopt a similar strategy to $\theta^\mu_{AA'}$. A practical way to do this is described in Appendix B.2 of \cite{Dreiner:2008tw}, with some changes in conventions. In particular, it is explained in this work that some of the soldering form identities do continue to hold even in $d\not=4$ dimensions. The particular identity that is of interest to us is
\be
\theta^{(\mu A}{}_{E'} \theta^{\nu) B E'} = -\frac{1}{2} \eta^{\mu\nu} \epsilon^{AB},
\ee
where now $\eta^{\mu\nu}$ is the $d$-dimensional metric, while the spinor indices $A,A'$ continue to be 2-dimensional. In particular $\epsilon_A{}^A=2$ as before. Contracting this identity with $k_\mu k_\nu$ we have
\be\label{k-dimreg}
k^A{}_{E'} k^{B E'} = - \frac{1}{2} \epsilon^{AB} k^2,
\ee
which is the identity we have used in proving (\ref{ident}), but which now holds for a $d$-dimensional momentum. 

To summarise, our discussion has established that the identity (\ref{k-dimreg}) can still be used even in dimensional regularisation. One can thus continue to use the Feynman rules in their spinor form derived above. One should just be careful not to use in manipulations with the loop momenta some identity of the soldering forms that only holds in four dimensions. In calculations below we shall only use the identity (\ref{k-dimreg}), which is also what is required for the property (\ref{ident}) to hold. 

It is important to stress that in our scheme the gauge field $A^a_{AA'}$ remains a 4-dimensional object, i.e. a spinor. In other words, we {\it do not} write $A^a_{AA'} = -\theta^\mu_{AA'} A_\mu^a$ and continue $A_\mu^a$ to $d$ dimensions. There is no need to do this in our setting as all Lagrangians and the resulting Feynman rules directly operate with objects $k_{AA'}$ and $A^a_{AA'}$. We need to continue $k_{AA'}$ to $d$ dimensions for reasons of regularising the divergent integrals, but there is no reason to do the same with $A^a_{AA'}$. The propagator can continue to be given by (\ref{propagator}), with the square of the $d$-dimensional $k_\mu$ in the denominator, but the product of the usual 2-dimensional spinor metrics in the numerator. This is essentially the same scheme that is used in the 2-component version of QED in \cite{Dreiner:2008tw}, with understanding that the field $A^a_{AA'}$ is now treated as a spinor, not tensor object. With these choices there is no difficulty in treating the considered here ``self-dual'' objects also in dimensional regularisation. 

\subsection{There are never two dots on any internal line}

Having established that our identity (\ref{ident}) also holds in dimensionally regularised loop integrals, we would like to analyse consequences of this for the power counting. The identity (\ref{ident}) says that when two derivatives happen to act on the same internal line, this collapses the line. Importantly, this is as general as we stated it, because the derivatives of the gauge field only appear in the Lagrangian in the combination (\ref{da}). In other words, they only appear as dots, and the property (\ref{ident}) applies. This means that there can never be more than one derivative on any internal line of any Feynman diagram that we need to analyse. We have already used this fact in our analysis of the large $z$ behaviour of amplitudes under the BCFW shifts. Here we will analyse consequences of this fact for the power counting. 

\subsection{Power-counting in effective field theory}

To understand the power counting, let us first recall the power-counting in a ``typical'' effective field theory. In this case we have Weinberg's formula \cite{Weinberg:1978kz}
\be\label{weinberg}
D = 2L + 2 + \sum_k (k-2) m_k,
\ee
where $D$ is the superficial degree of divergence, $L$ is the number of loops and $m_k$ is the number of vertices in the diagram with $k$ derivatives in them. In case of many effective field theories, such as e.g. the chiral perturbation theory or gravity, there can only be an even number of derivatives at each vertex, and this number is greater or equal than two. Then the last term in (\ref{weinberg}) is a sum of non-negative terms. A particular case is that of perturbative gravity, where all vertices have two derivatives in them. In this case $D_{\rm GR}=2L+2$. The relation (\ref{weinberg}) means that for each given $D$ there is only a finite number of diagrams to consider. The relation (\ref{weinberg}) is easily obtained from the two obvious relations 
\be
D=4L - 2I + \sum_k k \, m^k, \qquad L = I - \sum_k m^k + 1,
\ee
where $I$ is the number of internal lines. 

\subsection{Power-counting in gauge theory}

In the case of gauge theories, such as theories of type $f(F)$, with $F$ being the (full or self-dual) curvature, the formula (\ref{weinberg}) still applies, but one also has vertices with one or no derivatives in them. Then the last term in (\ref{weinberg}) for $k=0,1$ contributes negatively. In such a situation this formula is not very useful. And indeed, in the case of usual YM theory, when we have just the cubic single derivative vertex and quartic no-derivative vertex we have an additional relation $E+2I = 3V_3+4V_4$, which implies $D_{\rm YM}=4-E$ typical of a renormalisable theory. 

So, let us write down the power-counting relations for a gauge theory of the type $f(F)$, first without taking into account the collapsing property (\ref{ident}). It is easy to see that in a theory of such type, at a vertex of valency $n$ there can be as many as $n$ derivatives, as well as $n-2, n-4, \ldots$, i.e. the number of derivatives decreases by two. An example of this is the usual 1-derivative YM cubic vertex and the 3-derivative new vertex. Thus, we label the number of $n$-valent vertices with $k$ derivatives in them by $V_n^k$. We then have
\be\nonumber
D = 4L - 2I + \sum_{n=3}^\infty \sum_{i=0}^{[n/2]}  (n-2i)  V^{n-2i}_n, \\ \label{relations}
L = I - \sum_{n=3}^\infty \sum_{i=0}^{[n/2]}  V^{n-2i}_n + 1, \\ \nonumber
E + 2I = \sum_{n=3}^\infty \sum_{i=0}^{[n/2]} n V^{n-2i}_n,
\ee
where we indicated the summation limits explicitly. From the first and the last of these relations we get
\be
4L+E-D = \sum_{n=3}^\infty \sum_{i=0}^{[n/2]} (2i) V^{n-2i}_n \geq 0,
\ee
and thus the most divergent diagrams are those where all vertices have the maximal number of derivatives possible. In other words, we have 
\be\label{max-D}
D \leq 4L + E,
\ee
with the equality being realised when all vertices have the maximal number of derivatives possible. 

It also should be noted that the maximal possible degree according to (\ref{max-D}) is actually an overcount as far as the divergence is concerned. Indeed, to get the maximal possible degree of divergence we take all vertices to have the number of derivatives equal to their valency. It is then clear that precisely $E$ of the derivatives are located on the external lines, and do not contribute to the divergence. Thus, the formula $D_{\rm max}=4L+E$ should actually be interpreted as a single quartic divergence per every loop integration. 

The degree of divergence that increases with the number of loops as $4L$ is clearly much worse than the divergence increasing as $2L$, which is the case in e.g. GR. Below we will see that the collapsing property implies that the degree of divergence for our theories is in between these two cases. 

\subsection{Power counting for deformations of YM}

We now take into account the property that when two derivatives act on a single internal line, the line collapses. This property implies that the divergence is never as bad as in (\ref{max-D}). Indeed, to get an equality in (\ref{max-D}) one needs to cancel all propagators by numerators, which precisely means two derivatives on each internal line. 

Let us write down relations for the power counting in the case when the maximal number of derivatives on each line is one. Let us denote by $I_1$ the number of internal lines with a single derivative, and $I_0$ those with no derivatives. We then have
\be\label{D-us}
D= 4L - I_1 - 2I_0 +E_1,
\ee
where $E_1$ is the number of derivatives on the external lines. 

Let us now compute the largest possible degree of divergence given the number of loops and the number of external lines. It is clear from (\ref{D-us}) that the largest degree of divergence is achieved when all internal lines carry one derivative. So, let us assume that $I=I_1$. From the second relation in (\ref{relations}) we see that we minimise the number of internal lines at fixed $L$ by having as small number of vertices as possible. Since loop diagrams with just one vertex vanish in dimensional regularisation, we need at least two vertices, which means that 
\be
I_{\rm min} = L+1,
\ee
and
\be
D_{\rm max} = 3L-1+E_1.
\ee

It remains to compute the maximal possible number of the derivatives on the external lines. It may seem at first that this number is equal to the number of external lines, but this is not true. Indeed, consider for example the diagram
\be
 \begin{fmfgraph*}(40,30) 
             \fmfleft{i}
             \fmfright{o}
             \fmf{dotted}{v1,i}
           	   \fmf{dotted, tension=0.2, right}{v1,v2} 
           	   \fmf{dotted, tension=0.2, left}{v1,v2} 
	    \fmf{dotted, tension=0.2}{v1,v2} 
	     \fmf{dotted}{v2,o}
       \end{fmfgraph*}
    \ee
It satisfies all the requirements assumed above: minimal number of vertices and thus maximal number of internal lines, each internal line carries single derivative. However, there is no quartic vertex with single derivative that could be used in this diagram, and so we learn that the maximal value of $E_1$ cannot always be equal to $E$. The possible diagram in this case is
\be
 \begin{fmfgraph*}(40,30) 
             \fmfleft{i}
             \fmfright{o}
             \fmf{dotted}{v1,i}
           	   \fmf{dotted, tension=0.2, right}{v1,v2} 
           	   \fmf{dotted, tension=0.2, left}{v1,v2} 
	    \fmf{dotted, tension=0.2}{v1,v2} 
	     \fmf{wiggly}{v2,o}
       \end{fmfgraph*}
    \ee
and has the degree of divergence $D=2\cdot 4 - 3+1 = 6$.

Simple analysis then shows that in the situation of minimal $I$ the maximal number of dots on the external lines is $E$ when $L$ is odd, and $E-1$ when $L$ is even. Thus, we get
\be\label{D-max}
D_{\rm max} = \left\{ \begin{array}{c} 3L-1+E, \qquad L\,\, {\rm odd} \\ 3L-2+E, \qquad L\,\, {\rm even}
\end{array} \right.
\ee
In particular, this formula implies that $D_{\rm max}-E$ is always even. 

The formula (\ref{D-max}) should be compared to (\ref{max-D}) and (\ref{weinberg}). It is clear that the divergences in our case are better than (\ref{max-D}), because the degree of divergence increases only as $3L$ and not $4L$ as would be the case in general theories of type $f(F)$. On the other hand, we get a faster growth than e.g. in the case of perturbative gravity with $D=2L+2$. This is because there can be an arbitrary number of derivatives in a vertex in our case, as compared to just two in the case of the perturbative expansion of the Einstein-Hilbert action. 

\subsection{1-loop self-energy diagrams}

In order to understand if our class of theories has any hopes of being renormalisable, let us consider the simplest possible divergences arising at one loop. Since we already know how the divergences of YM theory manifest themselves, we only need to consider the modifications due to new interactions. 

Thus, let us consider the new self-energy diagrams that appear when the theory is deformed. We use dimensional regularisation, with the only subtlety being that the dimensionally continued objects $k_{AA'}$ satisfy only some of their properties in four dimensions. Above we have established that (\ref{k-dimreg}) can continue to be used. 

We do not consider tadpole-type diagrams that can be built from quartic vertices because these do not have any divergences in dimensional regularisation used here. This also applies to diagrams built from two cubic vertices where there is a pair of dots on one of the (or both) lines, because these diagrams collapse to tadpole-type ones. Thus, in particular there is no divergent diagram built from two vertices (\ref{v3new}), because both internal lines of this diagram collapse. 

The only diagram whose divergence is non-zero in dimensional regularisation is one built from the new cubic and the YM cubic vertices, and where the dot of the YM vertex is on the external line
  \be
  \begin{fmfgraph*}(40,20) 
             \fmfleft{i}
             \fmfright{o}
             \fmf{dotted}{v1,i}
           	   \fmf{dotted, tension=0.5, right}{v1,v2} 
           	   \fmf{dotted, tension=0.5, left}{v1,v2} 
               \fmf{dotted}{v2,o}
       \end{fmfgraph*}
    \ee
It is clear that this diagram diverges quadratically. Since there are already two derivatives present on the external lines, it seems that the divergent part of this diagram requires a counterterm of type that is not present in the original Lagrangian. This is true, but just to an extent, because this counterterm corresponds to a simple field redefinition, as we shall now see.

Using the Feynman rules, we see that to evaluate the above diagram we need to perform the Wick contraction in the following expression
\be
\frac{\alpha g}{2M^2} f^{abc} (k A(k)))^a_A{}^B (-l A(-l))^b_B{}^C ((l-k) A(l-k))^c_C{}^A \\ \nonumber
g f^{def} (-kA(-k))^{d MN} A^e_{MM'}(-l+k) A^f_N{}^{M'}(l),
\ee
where $1/2$ is the symmetry factor. The Wick contraction gives
\be\label{bub-1}
\frac{\alpha g^2}{2M^2}(-C_2) (k A(k)))^a_A{}^B (kA(-k))^{a MN} 2l_{(B A'} \epsilon^{C)}{}_N \epsilon^{A'M'} 2 (l-k)_{(CB'} \epsilon^{A)}{}_M \epsilon^{B'}{}_{M'} \frac{1}{\im l^2 \im(l-k)^2}.
\ee
We now replace $l=q+xk, l-k=q+(x-1)k$, and keep only the quadratic and zeroth order terms in the integration variable $q$. We also note that the primed indices of $l$ and $l-k$ are getting contracted. So, we will also need $q_{BA'} q_C{}^{A'} = (1/2) \epsilon_{CB} q^2$, which continues to hold even in $d$ dimensions, as we have established in (\ref{k-dimreg}). With these replacements, the complicated structure appearing in (\ref{bub-1}) becomes
\be
\frac{1}{2}(q^2 + x(x-1) k^2) ( 4\epsilon_M{}^A \epsilon_{BN} + \epsilon_B{}^A \epsilon_{MN}).
\ee
The last term here is irrelevant as it contracts into the quantity that is $AB$ and $MN$ symmetric. 
The divergent part of the integral with $q^2$ is two times 
the divergent part of the integral without $q^2$ in the numerator,
thus overall, after performing the $x$ integration we get for this diagram
\be
\frac{1}{\epsilon} \frac{k^2}{M^2} \frac{\im \alpha g^2 C_2}{(4\pi)^2} (k A(k)))^a_{AB} (k A(-k)))^{a AB} .
\ee
This means that this divergence is removed by the following local field redefinition
\be\label{field-redef}
A\to A + \frac{1}{\epsilon} \frac{\alpha g^2 C_2}{(4\pi)^2} \frac{k^2}{M^2} A
\ee
applied to the two point function $-(\im/2) (k A(k)))^a_{AB} (k A(-k)))^{a AB}$.

\subsection{General case}

Let us now try to generalise the previous example. We know that at the one loop level, the maximal possible degree of divergence of a diagram is $D_{\rm max}=2+E$. Since we already have in our Lagrangian $E$-valent vertices with $E$ derivatives in them, we see that there is a hope to renormalise the theory at one loop by removing the divergences of the maximal degree $D_{\rm max}=2+E$ using the field redefinition (\ref{field-redef}). Indeed, by applying this field redefinition to all the vertices of the original Lagrangian one may hope to remove all the maximal degree divergences. What remains are then the divergences of degree $E$, which are just the logarithmic divergences contributing to the multiplicative renormalisation of all the couplings. Below we will see how to realise this program at one loop for the renormalisation of the coupling constant $\alpha$. 

We did not work out any details for the higher loop case, but if renormalisation is to work, it has to proceed by the local field redefinition of the type
\be
A\to A + \sum_{n=1}^\infty C_n \left(\frac{k^2}{M^2}\right)^{n} A,
\ee
where $C_n$ are divergent coefficients, followed by multiplicative renormalisation of all the couplings. One may also need non-linearities in the gauge field appearing in the field redefinition. Little more can be said without doing explicit calculations at higher loops, and we leave this to future work. 

\section{$\beta$-function calculation}
\label{sec:beta}

In dimensional regularisation (unlike in any Wilsonian scheme) only dimensionless couplings run. In our case all dimensionful couplings are expressed as multiples of powers of the same dimensionful parameter $M$. In dimensional regularisation this scale cannot run, with only the dimensionless couplings sitting in front of inverse powers of $M$ receiving (infinite) dimensionless corrections. This will be explicitly visible in the computations we perform, in the sense that the inverse powers of $M$ present in the vertices always just go outside of the divergent integrals to be computed, and serve as the bookkeeping device for which term in the Lagrangian the divergence in question contributes to. To put it concisely, the scale $M$ does not run in dim reg, and only the dimensionless couplings do. 

This feature of the scheme we use is quite important, for it implies that the presence of the new couplings does not change the running of the YM coupling $g^2$. This decoupling of the flow of $g^2$ from the flow for the other couplings does not occur in Wilsonian schemes. We find this decoupling nice and highly desirable, which serves as one more motivation to prefer the dimensional regularisation to any other possible scheme. 

Our strategy for computing the $\beta$-function for the new coupling $\alpha$ is as follows. As we have seen previously, the presence of the new vertex does not change the multiplicative renormalisation of the gauge field from what it is in pure YM. Thus, we do not need to recompute the self-energy diagrams, as the result for the logarithmic divergence appearing in this case is known. We only need to consider the triangle diagrams. 

We then note that the presence of the new vertex cannot affect the logarithmic divergences contributing to the renormalisation of the usual YM vertex. This follows from the fact that the new vertex comes with a dimensionful 
prefactor of $1/M^2$. Thus, whatever the divergences that the diagrams with the new vertex produce, they require counterterms of different mass dimension. Thus, there cannot be contributions to the running of the YM coupling $g$ from the presence of the new interactions. 

The new interaction can thus only contribute to its own strength running. In order to find this in as simple way as possible, we evaluate the divergences on shell, for a configuration of three positive helicity gluons, in the same fashion as the new vertex was evaluated on-shell (\ref{3-plus}) in our discussion of the scattering amplitudes. We remind the reader that the three external momenta need to be continued to complex values to obtain a non-zero result. This on-shell evaluation simplifies the calculation and allows us to extract the divergent contributions to the new vertex rather easily. 

  \subsection{3-point triangle diagrams}
  
 Let us consider possible triangle contributions to the vertex correction. We only need to analyse the effect of the new interactions, assuming the usual YM story as known. We note that diagrams with three new vertices completely collapse and are thus zero in dimensional regularisation. Diagrams with two new vertices become bubble diagrams proportional to the square of one of the external momenta, and hence are zero on-shell. Therefore we need only to consider the diagrams with one new cubic vertex. In this case there are three possible ways for each of the old vertex to place its dot, and thus in total nine diagrams to consider. In one of these diagrams the dots of both old vertices are placed on the internal lines already containing the dots of the new vertex.  
\unitlength = 1mm
\be
 \parbox{40mm}{\begin{fmfgraph*}(30,30) 
     \fmftop{i}
     \fmfbottom{o1,o2}
     \fmf{dotted}{v1,i}
   	   \fmf{phantom, tension=0.5,tag=1}{v1,v2} 
	    \fmf{phantom,tension=0.5,tag=1}{v1,v3} 
	      \fmf{wiggly, tension=0.3}{v2,v3}
	     \fmf{wiggly}{v2,o1}
       \fmf{wiggly}{v3,o2}
       \fmfposition
	   \fmfipath{p[]}
\fmfiset{p1}{vpath1(__v1,__v3)} 
\fmfi{dottedf}{subpath (0,length(p1)/2) of p1} \fmfi{dottedf}{subpath (length(p1)/2,length(p1)) of p1}	  
\fmfiset{p2}{vpath1(__v1,__v2)} 
\fmfi{dottedf}{subpath (0,length(p2)/2) of p2}  
\fmfi{dottedf}{subpath (length(p2)/2,length(p2)) of p2}	 
   	       \end{fmfgraph*}}  	
\ee
This diagram collapses and is again zero in dimensional regularisation. The remaining diagrams are those where no line collapses
\unitlength = 1mm
   \be\label{tr}
    \parbox{40mm}{\begin{fmfgraph*}(30,30) 
     \fmftop{i}
     \fmfbottom{o1,o2}
     \fmf{dotted}{v1,i}
   	   \fmf{dotted, tension=0.5}{v1,v2} 
   	   \fmf{dotted, tension=0.5}{v1,v3} 
   	   \fmf{wiggly, tension=0.3}{v2,v3}
       \fmf{dotted}{v2,o1}
       \fmf{dotted}{v3,o2}
        \end{fmfgraph*}} +
          \parbox{40mm}{\begin{fmfgraph*}(30,30) 
      \fmftop{i}
      \fmfbottom{o1,o2}
      \fmf{dotted}{v1,i}
    	   \fmf{dotted, tension=0.5}{v1,v2} 
    	   \fmf{dotted, tension=0.5}{v1,v3} 
    	   \fmf{dotted, tension=0.3}{v3,v2}
        \fmf{dotted}{v2,o1}
        \fmf{wiggly}{v3,o2}
         \end{fmfgraph*}}+
          \parbox{40mm}{\begin{fmfgraph*}(30,30) 
      \fmftop{i}
      \fmfbottom{o1,o2}
      \fmf{dotted}{v1,i}
    	   \fmf{dotted, tension=0.5}{v1,v2} 
    	   \fmf{dotted, tension=0.5}{v1,v3} 
    	   \fmf{dotted, tension=0.3}{v2,v3}
        \fmf{wiggly}{v2,o1}
        \fmf{dotted}{v3,o2}
         \end{fmfgraph*}} 
       \ee 
as well as diagrams where a single line collapses
   \be\label{tr-coll}
     \parbox{40mm}{\begin{fmfgraph*}(30,30) 
     \fmftop{i}
     \fmfbottom{o1,o2}
     \fmf{dotted}{v1,i}
   	   \fmf{dotted, tension=0.5}{v1,v2} 
	    \fmf{phantom,tension=0.5,tag=1}{v1,v3} 
	      \fmf{wiggly, tension=0.3}{v2,v3}
	     \fmf{dotted}{v2,o1}
       \fmf{wiggly}{v3,o2}
       \fmfposition
	   \fmfipath{p[]}
\fmfiset{p1}{vpath1(__v1,__v3)} 
\fmfi{dottedf}{subpath (0,length(p1)/2) of p1} \fmfi{dottedf}{subpath (length(p1)/2,length(p1)) of p1}	    	       \end{fmfgraph*}} +
          \parbox{40mm}{\begin{fmfgraph*}(30,30) 
      \fmftop{i}
      \fmfbottom{o1,o2}
      \fmf{dotted}{v1,i}
    	   \fmf{phantom, tension=0.5,tag=1}{v1,v2} 
    	   \fmf{dotted, tension=0.5}{v1,v3} 
    	   \fmf{wiggly, tension=0.3}{v2,v3}
        \fmf{wiggly}{v2,o1}
        \fmf{dotted}{v3,o2}
          \fmfposition
	   \fmfipath{p[]}
\fmfiset{p1}{vpath1(__v1,__v2)} 
\fmfi{dottedf}{subpath (0,length(p1)/2) of p1} \fmfi{dottedf}{subpath (length(p1)/2,length(p1)) of p1}
         \end{fmfgraph*}} 
+
   \parbox{40mm}{\begin{fmfgraph*}(30,30) 
     \fmftop{i}
     \fmfbottom{o1,o2}
     \fmf{dotted}{v1,i}
   	   \fmf{dotted, tension=0.5}{v1,v2} 
	    \fmf{phantom,tension=0.5,tag=1}{v1,v3} 
	      \fmf{dotted, tension=0.3}{v2,v3}
	     \fmf{wiggly}{v2,o1}
       \fmf{wiggly}{v3,o2}
       \fmfposition
	   \fmfipath{p[]}
\fmfiset{p1}{vpath1(__v1,__v3)} 
\fmfi{dottedf}{subpath (0,length(p1)/2) of p1} \fmfi{dottedf}{subpath (length(p1)/2,length(p1)) of p1}	    	       \end{fmfgraph*}} +
          \parbox{40mm}{\begin{fmfgraph*}(30,30) 
      \fmftop{i}
      \fmfbottom{o1,o2}
      \fmf{dotted}{v1,i}
    	   \fmf{phantom, tension=0.5,tag=1}{v1,v2} 
    	   \fmf{dotted, tension=0.5}{v1,v3} 
    	   \fmf{dotted, tension=0.3}{v3,v2}
        \fmf{wiggly}{v2,o1}
        \fmf{wiggly}{v3,o2}
          \fmfposition
	   \fmfipath{p[]}
\fmfiset{p1}{vpath1(__v1,__v2)} 
\fmfi{dottedf}{subpath (0,length(p1)/2) of p1} \fmfi{dottedf}{subpath (length(p1)/2,length(p1)) of p1}
         \end{fmfgraph*}}           \ee 
 plus an additional  diagram
 \be\nonumber
 \parbox{40mm}{\begin{fmfgraph*}(30,30) 
      \fmftop{i}
      \fmfbottom{o1,o2}
      \fmf{dotted}{v1,i}
    	   \fmf{dotted, tension=0.5}{v1,v2} 
    	   \fmf{dotted, tension=0.5}{v1,v3} 
    	   \fmf{phantom, tension=0.3,tag=1}{v3,v2}
        \fmf{wiggly}{v2,o1}
        \fmf{wiggly}{v3,o2}
          \fmfposition
	   \fmfipath{p[]}
\fmfiset{p1}{vpath1(__v3,__v2)} 
\fmfi{dottedf}{subpath (0,length(p1)/2) of p1} \fmfi{dottedf}{subpath (length(p1)/2,length(p1)) of p1}
         \end{fmfgraph*}}     
               \ee
In this last diagram the bottom line collapses, and the result is a diagram proportional to $k^{2}$ and thus zero on shell. It remains to consider diagrams (\ref{tr}) and (\ref{tr-coll}). 

Let us first consider (\ref{tr-coll}). We need to compare the resulting collapsed diagrams to similar diagrams that appear if one pairs the old cubic and the new quartic vertex, see below. To evaluate the resulting collapsed diagrams we use the rule (\ref{2dotscontr}) which can be read as saying that 
whenever the two dots appear on the same internal line, the line collapses in the sense of being replaced by the identity matrix in the space of symmetric rank 2 spinors, times the numerical factor of $2\I$, which is what the numerical factor in (\ref{2dotscontr}) becomes if one takes into account an additional sign from the fact that one of the momenta is outgoing and the other incoming. 
Carefully comparing the corresponding Feynman rules we find 
\be
 \parbox{40mm}{\begin{fmfgraph*}(30,30) 
     \fmftop{i}
     \fmfbottom{o1,o2}
     \fmf{dotted,label=$k_{1}$,label.side=right}{v1,i}
   	   \fmf{dotted, tension=0.5}{v1,v2} 
	    \fmf{phantom,tension=0.5,tag=1}{v1,v3} 
	      \fmf{wiggly, tension=0.3}{v2,v3}
	     \fmf{dotted,label=$k_{3}$,label.side=right}{v2,o1}
       \fmf{wiggly,label=$k_{2}$,label.side=left}{v3,o2}
       \fmfposition
	   \fmfipath{p[]}
\fmfiset{p1}{vpath1(__v1,__v3)} 
\fmfi{dottedf}{subpath (0,length(p1)/2) of p1} \fmfi{dottedf}{subpath (length(p1)/2,length(p1)) of p1}	    	       \end{fmfgraph*}} = \qquad - \quad
\parbox{40mm}{\begin{fmfgraph*}(30,30) 
     \fmfleft{i}
     \fmfright{o1,o2}
     \fmf{dotted,label=$k_{3}$,label.side=right}{v1,i}
   	   \fmf{dotted, tension=0.5,right}{v2,v1} 
	    \fmf{wiggly,tension=0.5,left}{v2,v1}
	     \fmf{dotted,label=$k_{1}$,label.side=right}{v2,o2}
       \fmf{wiggly,label=$k_{2}$,label.side=left}{v2,o1}
          	       \end{fmfgraph*}}
\ee
and
\be
 \parbox{40mm}{\begin{fmfgraph*}(30,30) 
     \fmftop{i}
     \fmfbottom{o1,o2}
     \fmf{dotted,label=$k_{1}$,label.side=right}{v1,i}
   	   \fmf{dotted, tension=0.5}{v1,v2} 
	    \fmf{phantom,tension=0.5,tag=1}{v1,v3} 
	      \fmf{dotted, tension=0.3}{v2,v3}
	     \fmf{wiggly,label=$k_{3}$,label.side=right}{v2,o1}
       \fmf{wiggly,label=$k_{2}$,label.side=left}{v3,o2}
       \fmfposition
	   \fmfipath{p[]}
\fmfiset{p1}{vpath1(__v1,__v3)} 
\fmfi{dottedf}{subpath (0,length(p1)/2) of p1} \fmfi{dottedf}{subpath (length(p1)/2,length(p1)) of p1}	    	       \end{fmfgraph*}}= \qquad - \quad
\parbox{40mm}{\begin{fmfgraph*}(30,30) 
     \fmfleft{i}
     \fmfright{o1,o2}
     \fmf{wiggly,label=$k_{3}$,label.side=right}{v1,i}
   	   \fmf{dotted, tension=0.5,right}{v2,v1} 
	    \fmf{dotted,tension=0.5,right}{v1,v2}
	     \fmf{dotted,label=$k_{1}$,label.side=right}{v2,o2}
       \fmf{wiggly,label=$k_{2}$,label.side=left}{v2,o1}
          	       \end{fmfgraph*}}
	       \ee
as well as similar relations for the other collapsing line. 

We can now see that the diagrams in (\ref{tr-coll}) precisely cancel similar diagrams from the old cubic new quartic vertex diagrams of the next section. Indeed, the symmetry factor in the triangle diagram case is $1/2$. When the new vertex is on the line with the external momentum $k_1$ we get four collapsing diagrams. In  two of them the momentum on the old cubic vertex of the effective bubble diagram is $k_3$ and in the other two it is $k_2$. When we take a sum over where the new cubic vertex can get attached we get overall four effective old cubic new quartic type diagrams with the momentum $k_1$ attached to the old cubic vertex. They all come with the symmetry factor of $1/2$, and in two of them there is one dot inside the loop, and in the other two there are two dots. Overall we get precisely the same diagrams as appear in the last two terms in (\ref{bubble}), with the same numerical factors, but with an opposite sign so these diagrams cancel with each other.

The only triangles remaining are then the two types of diagrams in  (\ref{tr}). The first one is
\be
\lower0.6in\hbox{\includegraphics[width=1.8in]{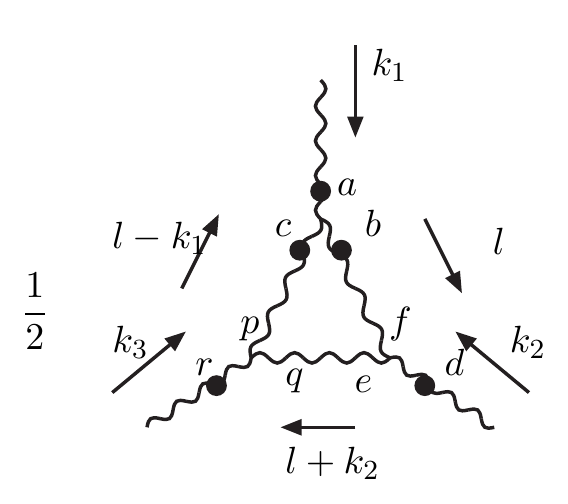}}
\begin{array}{ll}
= &  (\alpha g/2M^2) f^{abc}\Bigl(k_{1}A^{a}(k_{1})\Bigr)_{A}{}^{B}\Bigl(-lA^{b}(-l)\Bigr)_{B}{}^{C}\Bigl((l-k_{1})A^{c}(l-k_{1})\Bigr)_{C}{}^{A} \\
&  gf^{def}\Bigl(k_{2}A^{d}(k_{2})\Bigr)^{MN}A^{e}(-l-k_{2})_{MM'}A^{f}(l)_{N}{}^{M'} \\
& gf^{rpq}\Bigl(k_{3}A^{r}(k_{3})\Bigr)^{PQ}A^{p}(-l+k_{1})_{PP'}A^{q}(l+k_{2})_{Q}{}^{P'}
\end{array}
\ee
Wick contracting gives
\be \nonumber
\frac{\a g^{3}}{2M^2} \left(-\frac{C_{2}}{2}\right)f^{ade}\Bigl(k_{1}A^{a}(k_{1})\Bigr)_{A}{}^{B}\Bigl(k_{2}A^{d}(k_{2})\Bigr)^{MN}\Bigl(k_{3}A^{r}(k_{3})\Bigr)^{PQ}
\\ \nonumber \times 
2\, l_{(BB'}\epsilon^{C)}{}_{N}\,\epsilon^{B'M'}\,2(l-k_{1})_{(CC'}\epsilon^{A)}{}_{P}\,\epsilon^{C'}{}_{P'}\,\epsilon_{MQ}\,\epsilon_{M'}{}^{P'}
\, \frac{1}{\I l^2 \I (l-k_1)^2 \I (l+k_2)^2}.
\ee
Both the momenta in the numerator can be replaced by the loop integration momentum $q= l-x k_1 + y k_2$ because this is the only divergent contribution. After this replacement is done the complicated product of $\epsilon$'s and the $l,l-k_1$ momenta reduces to
\be\nonumber
2 q^2 \epsilon_{NB} \epsilon_P{}^A \epsilon_{QM},
\ee
plus a term proportional to $\eps_B{}^A$ which vanishes when contracted to the rest of the expression. The divergent part of this diagram is thus
\be\label{res1}
 \frac{1}{2} \frac{\a g^3 C_2}{(4\pi)^2M^2} \, \frac{2}{\epsilon} f^{abc} \Bigl(k_{1}A^{a}(k_{1})\Bigr)_{A}{}^{B}\Bigl(k_{2}A^{b}(k_{2})\Bigr)_{B}{}^{C}\Bigl(k_{3}A^{c}(k_{3})\Bigr)_{C}{}^{A}.
\ee

The other triangle diagram in  (\ref{tr}) that we need to compute is
\be
\lower0.6in\hbox{\includegraphics[width=1.8in]{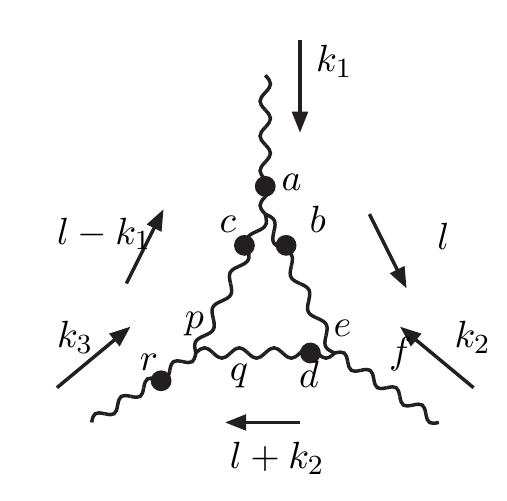}}
 \begin{array}{ll}
 = &  (\alpha g/M^2) f^{abc}\Bigl(k_{1}A^{a}(k_{1})\Bigr)_{A}{}^{B}\Bigl(-lA^{b}(-l)\Bigr)_{B}{}^{C}\Bigl((l-k_{1})A^{c}(l-k_{1})\Bigr)_{C}{}^{A} \\
 &  gf^{def}\left((-l-k_{2})A^{d}(-l-k_{2})\right)^{MN}A^{e}(l)_{MN'}A^{f}(k_{2})_{N}{}^{N'} \\
 &  gf^{rpq}\Bigl(k_{3}A^{r}(k_{3})\Bigr)^{PQ}A^{p}(-l+k_{1})_{PP'}A^{q}(l+k_{2})_{Q}{}^{P'}
 \end{array}
\ee
We do not include the symmetry factor of $1/2$ because there are two diagrams with the dot being applied to either the $k_3$ or $k_2$ legs. We shall see that on-shell the result is permutation symmetric, so adding the two diagrams cancels the symmetry factor.  Performing the contractions gives
\be
- \frac{\a g^3}{M^2} \left( -\frac{C_2}{2}\right) f^{afr} 
\Bigl(k_{1}A^{a}(k_{1})\Bigr)_{A}{}^{B}\, A^{f}(k_{2})_{N}{}^{N'}\,\Bigl(k_{3}A^{r}(k_{3})\Bigr)^{PQ}
\\ \nonumber \times
2l_{(BB'} \eps^{C)}{}_M \eps^{B'}{}_{N'} 2(l-k_1)_{(CC'} \eps^{A)}{}_P \eps^{C'}{}_{P'} 2 (l+k_2)^{(M}{}_{M'} \eps^{N)}{}_Q\eps^{M'P'}  \frac{1}{\I l^2 \I (l-k_1)^2 \I (l+k_2)^2},
\ee
where the minus in the brackets comes from the product of three structure constants and one extra overall minus comes from the fact that there are three propagators, each carrying a minus sign. We now need to open the symmetrisations, replacing the momenta with the loop integration momentum appropriately shifted, i.e. $l= q+xk_1-yk_2, l-k_1=q+(x-1)k_1-yk_2,l+k_2=q+xk_1+(1-y)k_2$. We then keep terms quadratic in $q$ and replace $q_\mu q_\nu$ with $(1/4)q^2$. 
Assuming the momenta $k_1,k_2$ are on-shell, $1_{AA'}=1_A 1_{A'}, 2_{AA'}=2_A 2_{A'}$,
and replacing the connections with the polarisation spinors using (\ref{kA}) gives the following result
\be\nonumber
4(1-3y) \ket{1}{2}\ket{2}{3}\ket{3}{1} + 2(1+3x) \ket{1}{2}\ket{3}{1}^2\frac{[p1]}{[p2]},
\ee
times a factor of $(-\im)^3$. Using the momentum conservation we can replace $[p1]/[p2]=-\ket{3}{2}/\ket{3}{1}$. Thus, after the integration over the Feynman parameters we get
\be\nonumber
(-\im)^3 4\ket{1}{2}\ket{2}{3}\ket{3}{1} = \frac{1}{2}\Bigl(k_{1}A(k_{1})\Bigr)_{A}{}^{B}\Bigl(k_{2}A(k_{2})\Bigr)_{B}{}^{C}\Bigl(k_{3}A(k_{3})\Bigr)_{C}{}^{A}.
\ee   
Thus, finally the divergent part of this diagram is
\be\label{res2}
-\frac{1}{4} \frac{\a g^3 C_2}{(4\pi)^2M^2} \, \frac{2}{\epsilon} f^{abc}\Bigl(k_{1}A^{a}(k_{1})\Bigr)_{A}{}^{B}\Bigl(k_{2}A^{b}(k_{2})\Bigr)_{B}{}^{C}\Bigl(k_{3}A^{c}(k_{3})\Bigr)_{C}{}^{A} .
\ee

\subsection{3-point bubble diagrams: New cubic-new quartic/New cubic-old quartic}

As for bubble diagrams, first we note that 
any dots collapsing occurring in a bubble diagram can only lead to tadpole or simply 
to a point, both are known to be vanishing. In the presence of a new cubic vertex 
this leaves us only the following two diagrams.
 \be  
        \parbox{40mm}{\begin{fmfgraph*}(30,30) 
      \fmftop{i}
      \fmfbottom{o1,o2}
      \fmf{dotted}{v1,i}
    	   \fmf{dotted,tension=0.5,right}{v1,v2} 
    	   \fmf{dotted,tension=0.5,left}{v1,v2} 
        \fmf{dotted}{v2,o1}
        \fmf{dotted}{v2,o2}
         \end{fmfgraph*}\,\,}  
 \hspace{2cm} 
       \parbox{40mm}{\begin{fmfgraph*}(30,30) 
      \fmftop{i}
      \fmfbottom{o1,o2}
      \fmf{dotted}{v1,i}
    	   \fmf{dotted,tension=0.5,right}{v1,v2} 
    	   \fmf{dotted,tension=0.5,left}{v1,v2} 
        \fmf{wiggly}{v2,o1}
        \fmf{wiggly}{v2,o2}
         \end{fmfgraph*}\,\,}  
   \ee
However a straightforward calculation shows that both diagrams are proportional to momentum square and hence are zero on-shell.

 \subsection{3-point bubble diagrams: Old cubic-new quartic}
There are three ways to attach the cubic vertex moving the dot and six ways to do the same with the quartic vertex. In total there are 18 diagrams. The symmetry factor for these diagrams is $1/4$ corresponding to flipping the two propagators, as well as flipping the external $23$ legs. Among the resulting 18 diagrams 6 vanish because there is a double dot appearing on a propagator. Among the remaining ones there are just 5 different diagram topologies to consider    
   \be     \label{bubble}
 \parbox{35mm}{\begin{fmfgraph*}(30,30) 
         \fmftop{i}
         \fmfbottom{o1,o2}
         \fmf{dotted}{v1,i}
       	   \fmf{wiggly,tension=0.5, right}{v1,v2} 
       	   \fmf{wiggly, tension=0.5, left}{v1,v2} 
           \fmf{dotted}{v2,o1}
           \fmf{dotted}{v2,o2}
            \end{fmfgraph*}\,\,}  + 
      2       \parbox{35mm}{\begin{fmfgraph*}(30,30) 
                   \fmftop{i}
                               \fmfbottom{o1,o2}
                   \fmf{wiggly}{v1,i}
                 	   \fmf{dotted,tension=0.5, right}{v1,v2} 
                 	   \fmf{wiggly, tension=0.5, left}{v1,v2} 
                     \fmf{dotted}{v2,o1}
                     \fmf{dotted}{v2,o2}
                 \end{fmfgraph*}\,\,}   +    
             4         \parbox{35mm}{\begin{fmfgraph*}(30,30) 
                   \fmftop{i}
                                     \fmfbottom{o1,o2}
                   \fmf{dotted}{v1,i}
                 	   \fmf{dotted,tension=0.5, left}{v2,v1} 
                 	   \fmf{wiggly, tension=0.5, left}{v1,v2} 
                     \fmf{wiggly}{v2,o1}
                     \fmf{dotted}{v2,o2}
                      \end{fmfgraph*}\,\,}  +
  4     \parbox{35mm}{\begin{fmfgraph*}(30,30) 
                         \fmftop{i}
                                                \fmfbottom{o1,o2}
                         \fmf{wiggly}{v1,i}
                       	   \fmf{dotted,tension=0.5, right}{v1,v2} 
                       	   \fmf{dotted, tension=0.5, right}{v2,v1} 
                           \fmf{wiggly}{v2,o1}
                           \fmf{dotted}{v2,o2}
                            \end{fmfgraph*}\,\,} 
                               \ee
For the last two diagrams it is understood that the four diagrams are actually $2+2$ with two different arrangements for where the dot goes on the bottom legs. One more diagram is
 \be
 \parbox{35mm}{\begin{fmfgraph*}(30,30) 
                      \fmftop{i}
                                              \fmfbottom{o1,o2}
                      \fmf{dotted}{v1,i}
                    	   \fmf{dotted,tension=0.5, left}{v2,v1} 
                    	   \fmf{dotted, tension=0.5, right}{v2,v1} 
                        \fmf{wiggly}{v2,o1}
                        \fmf{wiggly}{v2,o2}
                    \end{fmfgraph*}\,\,}
 \ee
 However, this last diagram is proportional to $k_1^2$ and hence zero on shell. It will not be considered any further. As we have seen above, the last two diagrams in (\ref{bubble}) are cancelled by what comes from the triangle diagrams. Hence, we only need to compute the first two diagrams.
   
The first one of them is very easy to compute 
\be\nonumber
\lower0.6in\hbox{\includegraphics[width=1.8in]{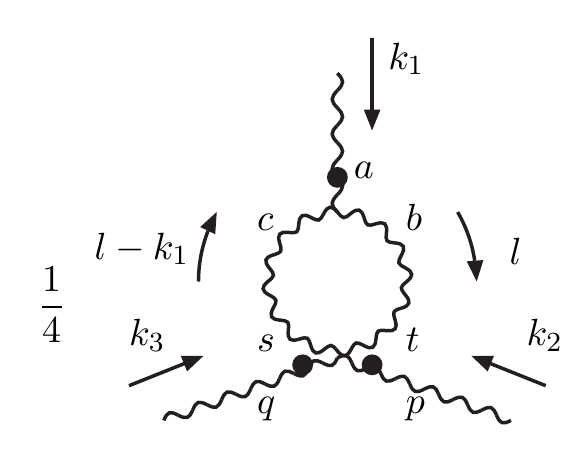}}
\begin{array}{ll}
= &  (1/4) gf^{abc}\Bigl(k_{1}A^{a}(k_{1})\Bigr)^{AB}A^{b}(-l)_{AB'}A^{c}(l-k_{1})_{B}{}^{B'} \\
& (-2i)\frac{\alpha g^{2}}{M^2} f^{rpq}f^{rst}\Bigl(k_{2}A^{p}(k_{2})\Bigr)_{M}{}^{N}\Bigl(k_{3}A^{q}(k_{3})\Bigr)_{N}{}^{P}A^{s}(-l+k_{1})_{(PP'}A^{t}(l)^{M)P'}
\end{array}
\ee
Here $1/4$ is the symmetry factor. We now do the contractions of the placeholders for the internal gauge field lines. This results in the following integrand
 \be     \nonumber
  \frac{1}{4} (-2\I) \frac{\a g^3}{M^2} (-C_2 f^{apq}) \Bigl(k_{1}A^{a}(k_{1})\Bigr)^{AB}\Bigl(k_{2}A^{p}(k_{2})\Bigr)_{M}{}^{N}\Bigl(k_{3}A^{q}(k_{3})\Bigr)_{N}{}^{P}
   \eps_{B(P}\eps^{B'}{}_{P'}\eps_{A}{}^{M)}\eps_{B'}{}^{P'} \frac{1}{\I l^2 \I (l-k_1)^2}.
   \ee
The divergent part of the loop integral is then
 \be\label{res3}
  -\frac{\a g^3 C_2}{(4\pi)^2M^2} \, \frac{2}{\epsilon} f^{abc}
  \Bigl(k_{1}A^{a}(k_{1})\Bigr)_{A}{}^{B}\Bigl(k_{2}A^{b}(k_{2})\Bigr)_{B}{}^{C}\Bigl(k_{3}A^{c}(k_{3})\Bigr)_{C}{}^{A}.
  \ee

The second diagram is a bit more difficult, but still rather straightforward
\be\nonumber
\lower0.6in\hbox{\includegraphics[width=1.8in]{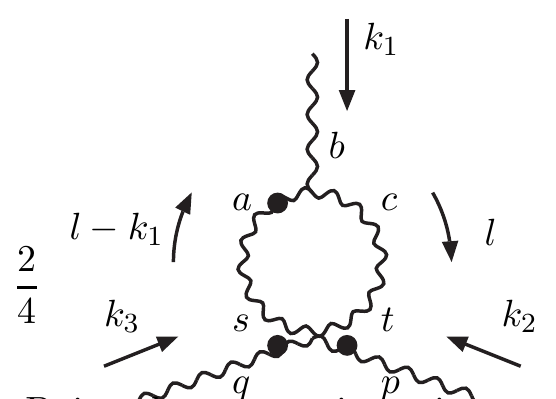}}= 
\begin{array}{ll}
& (1/4)gf^{abc}\Bigl((l-k_{1})A^{a}(l-k_{1})\Bigr)^{AB}A^{b}(k_{1})_{AB'}A^{c}(-l)_{B}{}^{B'}\\
& (-2i)\frac{\alpha g^{2}}{M^2}f^{rpq}f^{rst}\Bigl(k_{2}A^{p}(k_{2})\Bigr)_{M}{}^{N}\Bigl(k_{3}A^{q}(k_{3})\Bigr)_{N}{}^{P}A^{s}(-l+k_{1})_{(PP'}A^{t}(l)^{M)P'}
\end{array}
\ee
Doing the contractions gives
\be\nonumber
(-\I) \frac{\a g^3}{M^2} (-C_2)f^{bpq} \left( 
A^{b}(k_{1})_{AB'}\Bigl(k_{2}A^{p}(k_{2})\Bigr)_{M}{}^{N}\Bigl(k_{3}A^{q}(k_{3})\Bigr)_{N}{}^{P}
 2(l-k_1)^{(A}{}_{A'} \eps_{(P}{}^{B)} \eps_{P'}{}^{A'} \eps_{B}{}^{M)} \eps^{B'P'}   \right)
 \\ \nonumber \times \frac{1}{\I l^2 \I (l-k_1)^2}.
\ee
To evaluate this we use the fact when $l$ is replaced with $l=q+xk_1$ only the $k_1$ term survives. Thus, after some algebra of the $\epsilon$'s, and using the on-shell property of $A_1$, we get
 \be\nonumber
(\I)(x-1) \frac{\a g^3}{M^2} \;C_2 \;f^{bpq} \left( -\frac{1}{2} 
\Bigl(k_{1}A^{b}(k_{1})\Bigr)_{P}{}^{M}\Bigl(k_{2}A^{p}(k_{2})\Bigr)_{M}{}^{N}\Bigl(k_{3}A^{q}(k_{3})\Bigr)_{N}{}^{P}
\right) \frac{1}{\I l^2 \I (l-k_1)^2},
 \ee 
 with the divergent part evaluating to
  \be\label{res4}
 \frac{1}{4}\, \frac{\a g^3 C_2}{(4\pi)^2M^2} \, \frac{2}{\epsilon} f^{abc}\Bigl(k_{1}A^{a}(k_{1})\Bigr)_{A}{}^{B}\Bigl(k_{2}A^{b}(k_{2})\Bigr)_{B}{}^{C}\Bigl(k_{3}A^{c}(k_{3})\Bigr)_{C}{}^{A}.
  \ee

\subsection{$\beta$-function}
Finally, now we have all of the pieces required for the $\beta$-function calculation.
Combining the results of (\ref{res1}), (\ref{res2}), (\ref{res3}) and (\ref{res4}) we get
\be\nonumber
\left( \frac{1}{2} -\frac{1}{4} - 1 +\frac{1}{4}\right) = -\frac{1}{2}
\ee
and thus the divergent part of the 3-point function, evaluated on-shell is equal to
\be
- 6 \, \frac{\a g^3 C_2}{(4\pi)^2M^2} \, \frac{1}{\epsilon} f^{abc}
\Bigl(k_{1}A^{a}(k_{1})\Bigr)_{A}{}^{B}\Bigl(k_{2}A^{b}(k_{2})\Bigr)_{B}{}^{C}\Bigl(k_{3}A^{c}(k_{3})\Bigr)_{C}{}^{A},
  \ee
where we have multiplied the result by the $3!$ symmetry factor. In order to produce a finite result for the coupling, this divergence must be canceled by the following counterterm:
\be
Z_{\a g} = 1+ 6 \frac{g^2 C_2}{(4\pi)^2} \, \frac{1}{\epsilon}.
\label{zag}
\ee
Schematically, we can write the renormalised Lagrangian as
\be
-\frac{1}{2} Z_A (\partial A)^2 + Z_g g \mu^{\epsilon/2} A^2 (\partial A) + Z_{\a g} \mu^{\epsilon/2} \frac{\a g}{M^2} (\partial A)^3 + \ldots.
\ee
As remarked earlier, we absorb the self-energy bubble through a field redefinition, without writing the corresponding counterterm here. The mutiplicative renormalisation factor $Z_A$ is therefore unchanged and given by its YM value
\be
Z_A = 1+ \frac{10}{3} \frac{g^2 C_2}{(4\pi)^2} \, \frac{1}{\epsilon}.
\label{za}
\ee
In addition we have the following relations between the renormalised and bare quantities
\be
Z_A A^2 = A_0^2, \qquad Z_{\a g} \a g \mu^{\epsilon/2} A^3 = (\a g)_0 A_0^3.
\ee
From these two we get
\be
(\a g)_0 = \a g \mu^{\epsilon/2} Z_{\a g} Z_A^{-3/2}.
\ee
Combining the results from equations (\ref{zag}) and (\ref{za}) we have
\be
\log(Z_{\a g} Z_A^{-3/2}) = \frac{g^2 C_2}{(4\pi)^2} \, \frac{1}{\epsilon}
\ee
and therefore
\be
0=\frac{\partial (\a g)_0}{\partial \log(\mu)} = \frac{1}{\a g} \frac{\partial \a g}{\partial \log(\mu)} + \frac{\epsilon}{2}- \frac{g^2 C_2}{(4\pi)^2},
\ee
where we have used $\partial g^2 /\partial \log(\mu)= - \epsilon g^2 + \ldots$. We then arrive at a 
positive $\beta$-function 
\be\label{beta*}
 \frac{\partial \a g}{\partial \log(\mu)}  =  \frac{ \a g^3 C_2}{(4\pi)^2}.
 \ee

\subsection{Interpretation}

The result (\ref{beta*}) is our final result for the $\beta$-function of the new coupling. It nicely does not contain any numerical factors at all. As could have been anticipated, this sign of the $\beta$-function means that the new non-renormalisable interaction grows strong at high energies. 

We also note that the growth of $\alpha g$ is just logarithmic. Indeed, the coupling $\alpha g$ changes as $(g^2)^{-3/11}$. Given that the YM coupling changes with energy as
\be
g^2(\mu) = \frac{g^2(\mu_0)}{ 1+ (11 C_2/3(4\pi)^2) g^2(\mu_0) \log(\mu^2/\mu_0^2)}
\ee
we have
\be\label{running}
(\alpha g)(\mu) = (\alpha g)(\mu_0) \left( 1+ (11 C_2/3(4\pi)^2) g^2(\mu_0) \log(\mu^2/\mu_0^2)\right)^{3/11}.
\ee
Thus, if we deform the YM theory slightly by adding a new coupling $\a g$ at some energy scale, the departure from YM grows stronger as the energy increases. The flow will no longer take one to the simple free UV fixed point of YM theory. This can be taken as a manifestation of the asymptotic safety scenario: If one wants to end up at a given fixed point (free one in this case), one should stay on the critical surface which in this case means $\alpha g$ equal to zero. As the new coupling becomes order unity, the formula (\ref{running}) for its running can no longer be trusted. It is a very interesting question where the flow with non-zero $(\a g)_0$ takes one in the UV.

\section{Discussion}

In this paper we have motivated and introduced a new family of power-counting non-renormalisable gauge theories in four space-time dimensions. The Lagrangian is constructed from powers of the self-dual part of the field strength. This means that in the Lorentzian signature the Lagrangian is complex, which makes physical interpretation difficult, but we nevertheless proceeded with analysis of the scattering amplitudes and the renormalisation. 

For the scattering amplitudes our analysis can be justified by going to a (admittedly unphysical) setup of signature $(--,++)$, where the Lagrangian becomes real. In this case one can still talk about null momenta, and thus set up the scattering theory with its on-shell scattering amplitudes. Our first main result is then that the scattering amplitudes for our family of theories continue to be determinable from lower order amplitudes by BCFW recursion relations. For the case of amplitudes with at least one positive and one negative helicity gluon one can use the standard BCFW recursion with only a pair of momenta getting shifted. For the case of the all plus amplitudes one needs to use the more involved all-momentum shift. Another complication in the all plus amplitude case is that only a part of the amplitude is determined by the recursion. There is also a part of the $n$-point all plus amplitude that is obtained simply by evaluating the $n$-valent vertex on-shell. This way a number of new couplings enter the recursion at every particle number $n$. Thus, as we have already said in the Introduction, our theories are constructible in the sense that one does not need a Lagrangian to compute the amplitudes. They can all be determined from simple building blocks using the rules of recursion. 

We note that the all momentum shift has already been used in the context of on-shell reconstruction of general field theory amplitudes in \cite{Cohen:2010mi}. This work obtained a general condition for an applicability of the all-line shift to effective field theory amplitudes. Another related reference is \cite{Kampf:2012fn}, which considered the non-linear sigma-model. In this work the all-line shift was used to set up a BCFW recursion for the Berends-Giele current, from which the amplitudes can then be obtained by setting the single off-shell leg on-shell. 

It is clear that the described particular version of constructibility of our theories (in the sense that the amplitudes are determined from lower amplitudes by a recursion) relies on the special property (\ref{ident}) of the associated perturbation theory. It is this property that guaranteed the $1/z$ behaviour of the amplitudes under the BCFW and Risager's shifts. So, without this ``collapsing'' property many of the constructions of this paper are simply inapplicable. In particular, the usual 2-line BCFW shift is only available due to the property (\ref{ident}). However, the applicability of the Risager all-line shift is more general, as is explained in \cite{Cohen:2010mi}. Some amplitudes in a general effective field theory for the gauge field are constructible by the all-line shift; see \cite{Cohen:2010mi}.

We have also seen that the double copy structure of gravity amplitudes continues to hold even for "deformed" theories. Thus, we have seen that the graviton scattering amplitudes for "deformed" General Relativity continue to be given by squares of "deformed" YM amplitudes. This observation is not entirely new. In the context of parity-invariant gauge and gravity theories augmented by higher derivative operators it has been discussed in e.g. \cite{Cohen:2010mi}, and in much more details in \cite{Broedel:2012rc}. 

The usual explanation of the double copy pattern for gravity is the open/closed string duality. For some non-renormalisable operators this explanation is still valid, see \cite{Broedel:2012rc}. However, it appears unlikely that our chiral models with their infinite numbers of independent couplings can be realised in string theory. So, if the double copy pattern can be shown to hold in the context of theories with infinite number of independent couplings, then the open/closed string duality cannot be what is at the root of this pattern. Then the only other possible justification of the gravity equals YM squared behaviour would be the fact that the amplitudes in both cases are constructed using the same recursion relation from the building blocks that themselves follow the double copy rule. This has already been identified as the explanation of the double copy pattern in the case of usual GR and YM in \cite{BjerrumBohr:2010yc}; see also references therein. So, if the double copy pattern indeed extends to an infinite-parametric context as is suggested by our results, then one would be able to conclude that it is the constructible nature of the amplitudes that is at the heart of the corresponding double copy structure. 

As far as loop computations are concerned, we have seen that it is the property (\ref{ident}) that leads to an improved UV behaviour of our theories, in the sense that the degree of divergence grows slower than it would for a theory with as many as $n$ possible derivatives in an $n$-valent vertex. Given that in our case there is at most a single derivative on each internal Feynman diagram line, the divergences of our theories are in a sense similar to those of the usual YM theory. This does not mean that there are no divergences of a new type. But we have seen that at least at the one loop level, the new divergences can be absorbed into a simple local field redefinition. After this is done, there remain only logarithmic divergences contributing to the multiplicative renormalisation of the coupling constants. We then explicitly determined the resulting renormalisation group running of the coupling $\alpha g$, and found a positive $\beta$-function (\ref{beta*}). This is the second main result of this work. 

In the accompanying paper \cite{Krasnov:2015kva} we perform the complete one-loop analysis of our theories using the background field method. We will see that our expectation that at one loop the theory is renormalisable by a local field redefinition, as well as by multiplicative renormalisations, is realised. We will also explicitly compute the arising one loop beta-function. The result (\ref{beta*}) is then confirmed by a different calculation. 

Finally, let us discuss possible directions for future research. The most interesting question is whether the described here family of theories continues to be closed under renormalisation at higher loops. It is difficult to answer this without doing explicit calculations, and we leave this to the future. 

A related remark is that the open issues of reality of the Lagrangian have little to do with questions of renormalisability. Indeed, renormalisation is customarily studied in the Euclidean signature setup. In our case this would render the Lagrangian real, with the Euclidean path internal thus being of a non-oscillatory type. Thus, one can study divergences and renormalisation even without understanding issues related to the Lagrangian becoming complex for the Lorentzian signature. 

Another important question is whether one can make sense of the theory in the Lorentzian signature, by finding a positive-definite inner product that would make the theory unitary. This is a difficult question, probably requiring a better yet understanding of the finite-dimensional examples of the PT-symmetric quantum mechanics. 

\subsection*{Note added: Supersymmetry}

As suggested by an anonymous referee of this paper, we add a note addressing the question of whether it is possible to supersymmetrise our class of theories (\ref{L*}). We will only consider the case of $N=1$ supersymmetry.

In four space-time dimensions, $N=1$ super-YM theory is a theory of a gauge-field as well as a Lie algebra valued Majorana spinor. In 2-component spinor notations, a Majorana spinor is the Dirac spinor composed of a 2-component unprimed spinor and its complex conjugate primed one. The generator of supersymmetry is another Majorana spinor. 

Given that our approach is chiral, with one type of spinors playing a more important role than the other, and also that the requirement of Hermiticity of the Lagrangian is dropped, we can consider the following version of $N=1$ supersymmetric YM. The Lagrangian is given by
\be\label{susy-L}
{\cal L}= -\frac{1}{4} (F^a_{AB})^2 + \frac{1}{2} \bar{\psi}^{aA'} D_{A'}{}^{A} \psi^a_{A}.
\ee
The supersymmetry transformations are as follows
\be\label{susy}
\delta A_{AA'}^a = \bar{\psi}^a_{A'} \epsilon_A, \qquad \delta \psi^a_A = F^a_{AB} \epsilon^B.
\ee
Here $\epsilon^A$ is a (constant) SUSY generator. Note that in the above version of SUSY, only the unprimed Lie algebra valued spinor is transformed, with the primed one not changing under the supersymmetry transformation. Thus, here the unprimed $\psi^a_{A'}$ and primed $\psi^a_A$ spinors are not complex conjugates of each other, and the Lagrangian (\ref{susy-L}) is not Hermitian. Using $\delta F^{a AB}= D_{A'}{}^{A}  \delta A^{a A' B}$ and integrating by parts in the first term, it is easy to see that the Lagrangian (\ref{susy-L}) is invariant under (\ref{susy}), modulo a surface term. The term cubic in the Lie algebra valued spinor that results from the variation of the covariant derivative in the second term in (\ref{susy-L}) vanishes due to the Grassmann nature of the spinors $\bar{\psi}^a_{A'}$. 

Let us now consider a more general Lagrangian of the type studied in this paper, still supplemented with a kinetic term for a Lie algebra valued spinor
\be\label{L-susy-mod}
{\cal L} = f(F^a_{AB}) + \bar{\psi}^{aA'} D_{A'}{}^{A}  \psi^a_{A}.
\ee
Let us consider the supersymmetry variation of the first term in this Lagrangian, with the transformation rule for the gauge field unchanged from the one in (\ref{susy}). We get
\be
\delta f(F^a_{AB}) = (f')^a_{AB} D_{A'}{}^{A}  (\bar{\psi}^{aA'} \epsilon^B),
\ee
where $(f')^a_{AB}$ is the first derivative of the function $f$ with respect to its argument. It is then easy to see that to make (\ref{L-susy-mod}) SUSY-invariant we just have to change the definition of the transformation rule for the spinor $\psi^a_A$
\be
\delta \psi^a_A = (f')^a_{AB} \epsilon^B.
\ee
Thus, the bottom line is that our theories are as supersymmetrisable as the usual Yang-Mills, with only the transformation rule for the spinor $\psi^a_A$ needing a modification. 

\section*{Acknowledgements} The authors were supported by ERC Starting Grant 277570-DIGT. KK is grateful to Edward Witten for an email exchange regarding unitarity.

 \end{fmffile}


\begin{thebibliography}{99}

\bibitem{Weinberg:2009bg} 
  S.~Weinberg,
  ``Effective Field Theory, Past and Future,''
  PoS CD {\bf 09}, 001 (2009)
  [arXiv:0908.1964 [hep-th]].
  
\bibitem{Einhorn:2001kj} 
  M.~B.~Einhorn and J.~Wudka,
  ``Effective beta functions for effective field theory,''
  JHEP {\bf 0108}, 025 (2001)
  [hep-ph/0105035].
  
\bibitem{Krasnov:2015kva} 
  K.~Krasnov,
  ``One-loop $\beta$-function for an infinite-parameter family of gauge theories,''
  JHEP {\bf 1503}, 030 (2015)
  [arXiv:1501.00849 [hep-th]].
  
\bibitem{Heisenberg:1935qt} 
  W.~Heisenberg and H.~Euler,
  ``Consequences of Dirac's theory of positrons,''
  Z.\ Phys.\  {\bf 98}, 714 (1936); an English
translation available as physics/0605038.
 
\bibitem{Dixon:2004za} 
  L.~J.~Dixon, E.~W.~N.~Glover and V.~V.~Khoze,
  ``MHV rules for Higgs plus multi-gluon amplitudes,''
  JHEP {\bf 0412}, 015 (2004)
  [hep-th/0411092].

\bibitem{Britto:2005fq} 
  R.~Britto, F.~Cachazo, B.~Feng and E.~Witten,
  ``Direct proof of tree-level recursion relation in Yang-Mills theory,''
  Phys.\ Rev.\ Lett.\  {\bf 94}, 181602 (2005)
  [hep-th/0501052].
  
\bibitem{Krasnov:2006du} 
  K.~Krasnov,
  ``Renormalizable Non-Metric Quantum Gravity?,''
  hep-th/0611182.
  
\bibitem{Krasnov:2011up} 
  K.~Krasnov,
  ``Gravity as a diffeomorphism invariant gauge theory,''
  Phys.\ Rev.\ D {\bf 84}, 024034 (2011)
  [arXiv:1101.4788 [hep-th]].
  
\bibitem{Krasnov:2011pp} 
  K.~Krasnov,
  ``Pure Connection Action Principle for General Relativity,''
  Phys.\ Rev.\ Lett.\  {\bf 106}, 251103 (2011)
  [arXiv:1103.4498 [gr-qc]].
  
\bibitem{Delfino:2012aj} 
  G.~Delfino, K.~Krasnov and C.~Scarinci,
  ``Pure connection formalism for gravity: Feynman rules and the graviton-graviton scattering,''
  arXiv:1210.6215 [hep-th].
 
\bibitem{Broedel:2012rc} 
  J.~Broedel and L.~J.~Dixon,
  ``Color-kinematics duality and double-copy construction for amplitudes from higher-dimension operators,''
  JHEP {\bf 1210}, 091 (2012)
  [arXiv:1208.0876 [hep-th]].
 
\bibitem{Bender:2007nj} 
  C.~M.~Bender,
  ``Making sense of non-Hermitian Hamiltonians,''
  Rept.\ Prog.\ Phys.\  {\bf 70}, 947 (2007)
  [hep-th/0703096 [HEP-TH]].
 
\bibitem{Mostafazadeh:2010yx} 
  A.~Mostafazadeh,
  ``Conceptual Aspects of PT-Symmetry and Pseudo-Hermiticity: A status report,''
  Phys.\ Scripta {\bf 82}, 038110 (2010)
  [arXiv:1008.4680 [quant-ph]].
  
\bibitem{Mostafazadeh:2004qh} 
  A.~Mostafazadeh,
  ``PT-symmetric cubic anharmonic oscillator as a physical model,''
  J.\ Phys.\ A {\bf 38}, 6557 (2005)
  [Erratum-ibid.\ A {\bf 38}, 8185 (2005)]
  [quant-ph/0411137].
  
\bibitem{Boels:2008ef} 
  R.~Boels and C.~Schwinn,
  ``Deriving CSW rules for massive scalar legs and pure Yang-Mills loops,''
  JHEP {\bf 0807}, 007 (2008)
  [arXiv:0805.1197 [hep-th]].
  
\bibitem{Boels:2008fc} 
  R.~Boels, K.~J.~Larsen, N.~A.~Obers and M.~Vonk,
  ``MHV, CSW and BCFW: Field theory structures in string theory amplitudes,''
  JHEP {\bf 0811}, 015 (2008)
  [arXiv:0808.2598 [hep-th]].
  
 
\bibitem{Srednicki:2007qs} 
  M.~Srednicki,
  ``Quantum field theory,''
  Cambridge, UK: Univ. Pr. (2007) 641 p
  

  
\bibitem{ArkaniHamed:2008yf} 
  N.~Arkani-Hamed and J.~Kaplan,
  ``On Tree Amplitudes in Gauge Theory and Gravity,''
  JHEP {\bf 0804}, 076 (2008)
  [arXiv:0801.2385 [hep-th]].

\bibitem{Risager:2005vk} 
  K.~Risager,
  ``A Direct proof of the CSW rules,''
  JHEP {\bf 0512}, 003 (2005)
  [hep-th/0508206].
  
\bibitem{Cohen:2010mi} 
  T.~Cohen, H.~Elvang and M.~Kiermaier,
  ``On-shell constructibility of tree amplitudes in general field theories,''
  JHEP {\bf 1104}, 053 (2011)
  [arXiv:1010.0257 [hep-th]].
  
\bibitem{Bern:2002kj} 
  Z.~Bern,
  ``Perturbative quantum gravity and its relation to gauge theory,''
  Living Rev.\ Rel.\  {\bf 5}, 5 (2002)
  [gr-qc/0206071].
  
\bibitem{Dreiner:2008tw} 
  H.~K.~Dreiner, H.~E.~Haber and S.~P.~Martin,
  ``Two-component spinor techniques and Feynman rules for quantum field theory and supersymmetry,''
  Phys.\ Rept.\  {\bf 494}, 1 (2010)
  [arXiv:0812.1594 [hep-ph]].

\bibitem{Weinberg:1978kz} 
  S.~Weinberg,
  ``Phenomenological Lagrangians,''
  Physica A {\bf 96}, 327 (1979).

\bibitem{Kampf:2012fn} 
  K.~Kampf, J.~Novotny and J.~Trnka,
  ``Recursion Relations for Tree-level Amplitudes in the SU(N) Non-linear Sigma Model,''
  Phys.\ Rev.\ D {\bf 87}, 081701 (2013)
  [arXiv:1212.5224 [hep-th]].
  
\bibitem{BjerrumBohr:2010yc} 
  N.~E.~J.~Bjerrum-Bohr, P.~H.~Damgaard, B.~Feng and T.~Sondergaard,
  ``Proof of Gravity and Yang-Mills Amplitude Relations,''
  JHEP {\bf 1009}, 067 (2010)
  [arXiv:1007.3111 [hep-th]].

\end{thebibliography}
\end{document}